\documentclass[pageno]{jpaper}

\usepackage{times}            
\usepackage{helvet} 
\usepackage{url}                  
\usepackage{listings}          
\usepackage{enumitem}      
\usepackage{hyperref}   
\usepackage{graphicx}
\usepackage{mathpartir}	
\usepackage{ifsym}
\usepackage{subfig}
\usepackage{multirow}
\usepackage{amsmath}
\usepackage{epsfig}
\usepackage{amsfonts}
\usepackage{amssymb}

\usepackage[normalem]{ulem}

\renewcommand{\dots}{\ensuremath{...}}
\newcommand{\judgespace}{\hspace{0.2in}}
\newcommand{\securesql}{SecureSQL}

\newcommand{\LSQL}{\ensuremath{\lambda_{SQL}}}
\newcommand{\val}{\ensuremath{x}}
\newcommand{\expr}{\ensuremath{e}}
\newcommand{\stmt}{\ensuremath{s}}
\newcommand{\exprsub}[1]{\ensuremath{e_{#1}}}
\newcommand{\var}{\ensuremath{v}}
\newcommand{\func}{\ensuremath{f}}

\newcommand{\paramnamedrec}[1]{\ensuremath{[\name{1}:#1_{1},\dots,\name{m}:#1_{m}]}}
\newcommand{\namedrec}{\paramnamedrec{\expr}}

\newcommand{\tabl}[1]{\ensuremath{T_{#1}}}

\newcommand{\column}[1]{\ensuremath{c_{#1}}}
\newcommand{\tablexpr}[1]{\ensuremath{t_{#1}}}
\newcommand{\selectexpr}{\ensuremath{\sigma_{\exprsub{1}}(\exprsub{2})}}
\newcommand{\collist}{{\name{1}, \dots, \name{m}}}
\newcommand{\projectexpr}{\ensuremath{\pi_{\collist}(\expr{})}}
\newcommand{\productexpr}{\ensuremath{\expr{1}\times\expr{2}}}
\newcommand{\unionexpr}{\ensuremath{\exprsub{1}\cup\exprsub{2}}}
\newcommand{\diffexpr}{\ensuremath{\exprsub{1}\backslash\exprsub{2}}}

\newcommand{\ifexpr}{{\sf if}\ \exprsub{1}\ \stmt_{1}\ \stmt_{2}}
\newcommand{\ndencrypt}{r\_encrypt}
\newcommand{\nddecrypt}{r\_decrypt}
\newcommand{\aencrypt}{a\_encrypt}
\newcommand{\adecrypt}{a\_decrypt}
\newcommand{\dencrypt}{d\_encrypt}
\newcommand{\ddecrypt}{d\_decrypt}
\newcommand{\opencrypt}{op\_encrypt}
\newcommand{\opdecrypt}{op\_decrypt}
\newcommand{\key}{\ensuremath{k}}

\newcommand{\ins}{\ensuremath{\mathsf{insert}}}
\newcommand{\update}{\ensuremath{\mathsf{update}}}
\newcommand{\delete}{\ensuremath{\mathsf{delete}}}
\newcommand{\select}{\ensuremath{\mathsf{select}}}
\newcommand{\insertexpr}{\ensuremath{\ins_{\collist}(\tablexpr{})\ \expr{}}}
\newcommand{\updateexpr}{\ensuremath{\update_{\collist}(\tablexpr{})\ \expr{}}}
\newcommand{\deleteexpr}{\ensuremath{\delete(\tablexpr{})\ \expr{}}}
\newcommand{\selectstmt}{\ensuremath{\select_{\collist}(\expr{})}}

\newcommand{\typevar}[1]{\ensuremath{\tau_{#1}}}

\newcommand{\spt}{{\ensuremath{\sf CT}}}
\newcommand{\snde}{{\ensuremath{\sf RE}}}
\newcommand{\sade}{{\ensuremath{\sf AE}}}
\newcommand{\sde}{{\ensuremath{\sf DE}}}
\newcommand{\sope}{{\ensuremath{\sf OPE}}}
\newcommand{\void}{{\ensuremath{\sf Void}}}

\newcommand{\oppt}{{\ensuremath{\sf CT_{OPE}}}}
\newcommand{\dpt}{{\ensuremath{\sf CT_{DE}}}}
\newcommand{\ndpt}{{\ensuremath{\sf CT_{RE}}}}

\newcommand{\sndel}{{\ensuremath{\sf RE_{lval}}}}
\newcommand{\sadel}{{\ensuremath{\sf AE_{lval}}}}
\newcommand{\sdel}{{\ensuremath{\sf DE_{lval}}}}
\newcommand{\sopel}{{\ensuremath{\sf OPE_{lval}}}}

\newcommand{\rectype}{\ensuremath{[\typevar{1}, \dots, \typevar{n}]}}
\newcommand{\name}[1]{\ensuremath{n_{#1}}}
\newcommand{\namedrectype}{\ensuremath{[\name{1}:\typevar{1}, \dots, \name{n}:\typevar{n}]}}
\newcommand{\subtyperel}{\ensuremath{<:}}

\newcommand{\acylsubtyperel}{\ensuremath{<:}}
\newcommand{\coersionfunc}{\ensuremath{\sigma}}

\newcommand{\constraints}{\ensuremath{\Sigma}}

\newcommand{\typeenv}{\ensuremath{\Gamma}}
\newcommand{\satisfies}{\ensuremath{\vdash}}
\newcommand{\subs}{\ensuremath{\mathrm{S}}}
\newcommand{\unify}{{\sf UNIFY}}
\newcommand{\project}{{\sf PROJECT}}
\newcommand{\lvaltype}{{\sf LVALTYPE}}
\newcommand{\fresh}{{\sf FRESH()}}

\newcommand{\judgeexpr}[1]{\constraints_{#1},\typeenv_{#1}\satisfies\exprsub{#1} : \beta_{#1}}
\newcommand{\judgestmt}[1]{\constraints_{#1},\typeenv_{#1},\gamma_{#1}\satisfies\stmt_{#1}}

\newcommand{\policy}{\ensuremath{{E}}}

\newcommand{\prog}{\ensuremath{p}}

\newcommand{\phistory}[3]{\ensuremath{{H}}}

\newtheorem{theorem}{Theorem}[section]

\newtheorem{definition}[theorem]{Definition}

\newcommand{\card}[1]{\ensuremath{\mid\!\!{#1}\!\mid}}
\renewcommand{\paragraph}[1]{\noindent\textbf{#1}}

\begin{document}
\title{Information Flows in Encrypted Databases}
\author{
	Kapil Vaswani\\
	kapilv@microsoft.com
	\and
	Ravi Ramamurthy\\
	ravirama@microsoft.com
	\and
	Ramarathnam Venkatesan\\
	venkie@microsoft.com
}
\date{}
\maketitle
\thispagestyle{empty}
\begin{abstract}
In encrypted databases, sensitive data is protected from an untrusted server by encrypting columns using partially homomorphic encryption schemes, and storing encryption keys in a {\em trusted client}. 
However, encrypting columns and protecting encryption keys does not ensure confidentiality - sensitive data can leak during query processing due to information flows through the trusted client. 
In this paper, we propose \securesql, an encrypted database that partitions query processing between an untrusted server and a trusted client while ensuring the absence of information flows. 
Our evaluation based on OLTP benchmarks suggests that \securesql\ can protect against explicit flows with low overheads ($< 30\%$).
However, protecting against implicit flows can be expensive because it precludes the use of key databases optimizations and introduces additional round trips between client and server. 
\end{abstract}
\newcommand{\creditvar}{{\tt @c\_credit}}
\newcommand{\creditcol}{{\tt c\_credit}}

\section{Introduction}
The old adage that a chain is only as strong as its weakest link describes the state of data security in public cloud platforms. 
While encryption can protect data in cloud storage and during transit to/from the cloud, data appears in cleartext in main memory of untrusted servers during processing.
In the absence of defenses like the firewall, this window of vulnerability is an alluring target for malicious cloud administrators and malware. 
It is therefore not surprising that applications which handle sensitive information are usually not deployed on public cloud platforms.\\

\paragraph{Trusted clients.}
In the absence of practical schemes for fully homomorphic encryption~\cite{daniele10holygrail}, we consider a recently proposed computational model for ensuring data confidentiality based on encrypted databases and \textit{trusted clients}~\cite{hacigumus02sqlencrypted,popa2011cryptdb,tu2013processing}.
In encrypted databases, sensitive data is protected from an untrusted server by encrypting columns using partially homomorphic encryption (PHE) schemes, and storing encryption keys in a {\em trusted client}. 
For example, deterministic encryption schemes (e.g. AES in ECB  or CBC mode with fixed initialization vector~\cite{aesmodes2014}) permit equality checks on encrypted data.
Therefore, operations such as equi-joins, groupings, and set union/intersection can be performed without requiring encryption keys. 
Similarly, the Paillier cryptosystem supports addition of encrypted values. 
All other operations for which efficient homomorphic schemes are not known are delegated to a trusted client, a special node hosted in a trusted environment, potentially outside the public cloud.
The trusted client has access to encryption keys and performs computations that cannot be performed on encrypted data.

\paragraph{Information flows.}
Prior work (including CryptDB~\cite{popa2011cryptdb}) raises the vision of building an encrypted database using trusted clients and PHE. 
However, a key problem with this model is the lack of a simple, strong security property that can be enforced without a significant loss in performance. 
In prior work, security is based on the premise that the server does not have access to encryption keys. 
{\it However, protecting encryption keys does not guarantee confidentiality}.
Prior research~\cite{cedric2011information} shows that confidentiality can be achieved using a combination of semantically secure, randomized encryption~\cite{shafi1982probablistic} and information flow security. 
Unfortunately, in an encrypted database, randomized encryption negates critical optimizations such as indexes, and forces the database to offload almost all computation to the trusted client. 
Therefore, selective use of weaker encryption schemes (such as deterministic or order-preserving encryption) is unavoidable. 

However, using weaker encryption schemes not only weakens security of columns encrypted using those schemes, but it can also reveal information about other columns due to information flows. 
Consider a simple query {\tt INSERT INTO T (B) SELECT A FROM T}, where \textit{A} and \textit{B} are encrypted using randomized and deterministic encryption schemes respectively. 
In the trusted client model, this query can be executed by retrieving values from column \textit{A} from the server, decrypting them on the client, encrypting the values using a deterministic encryption algorithm, and writing the values to column \textit{B}. 
Notice that this query leaks the relative frequency of values in column A even though the keys are always protected! (See section~\ref{sec:problem} for more subtle examples). 
The threat models defined in prior work~\cite{popa2011cryptdb,tu2013processing} do not consider such information flows.

In this paper, we explore this security-performance trade-off. 
We define a security property that prevents such leaks while permitting the use of weaker encryption schemes. 
Informally, our security property prohibits \textit{information flows} from an encrypted column to other columns encrypted using weaker encryption schemes. 
This property forms a strong contract between the database and the developer. 
For example, if this property is enforced, the query described above will be not be permitted unless both columns are encrypted using sufficiently strong encryption schemes.  

\begin{figure}[t]
\includegraphics[scale=0.9]{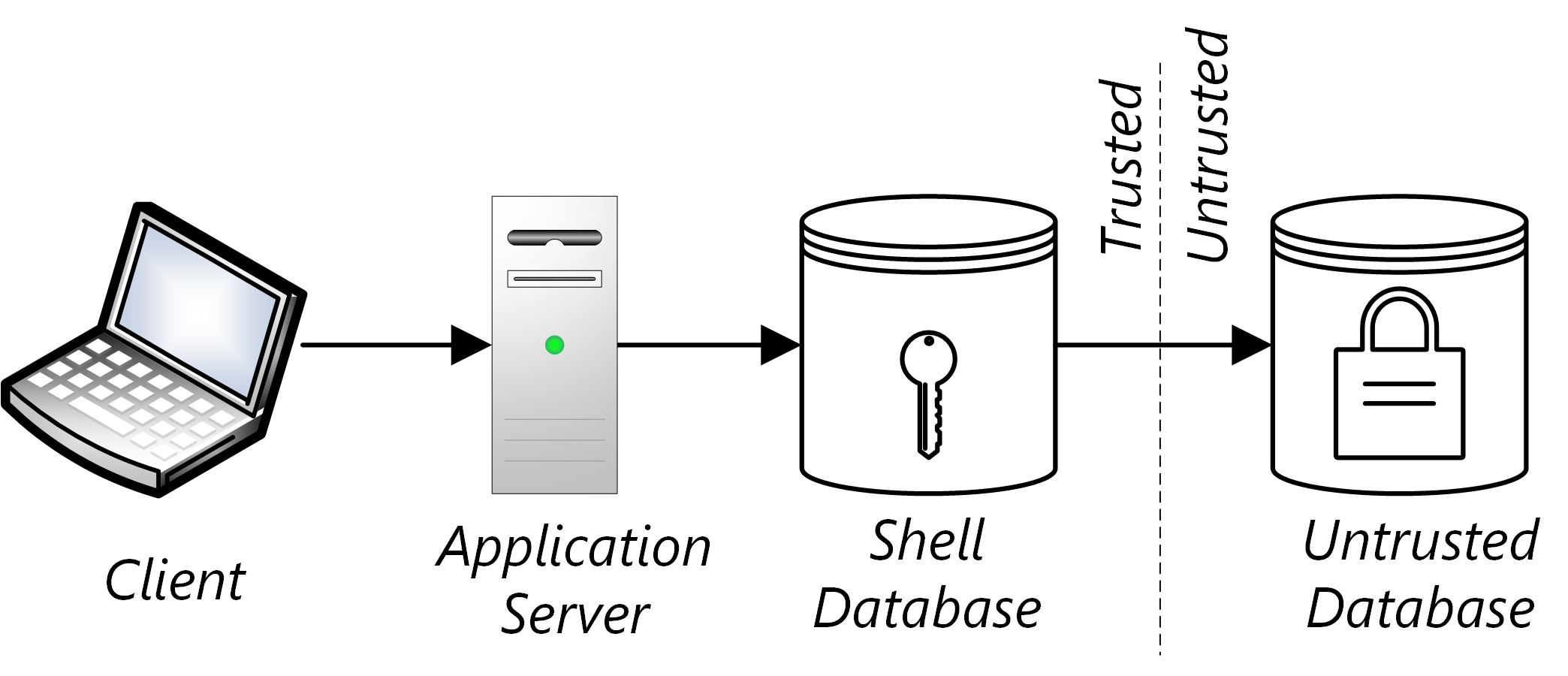}
\caption{
\label{fig:architecture}
Architecture for databases that use partially homomorphic encryptions and trusted client for data confidentiality}
\end{figure}

Next, we describe the design of \securesql, an encrypted database which enforces this property. 
Figure~\ref{fig:architecture} shows the architecture of \securesql. 
In \securesql\, the trusted client is a "empty" database (referred to as the \textit{shell database}) which stores encryption keys and performs residual query processing. 
Applications connect and run queries against the shell database, which orchestrates query processing with the server (using a commonly available database abstraction for distributed query processing known as \textit{linked server}~\cite{linkedserver13}) and returns cleartext results to the application. 

\begin{figure}[t]
\includegraphics[width=3.25in]{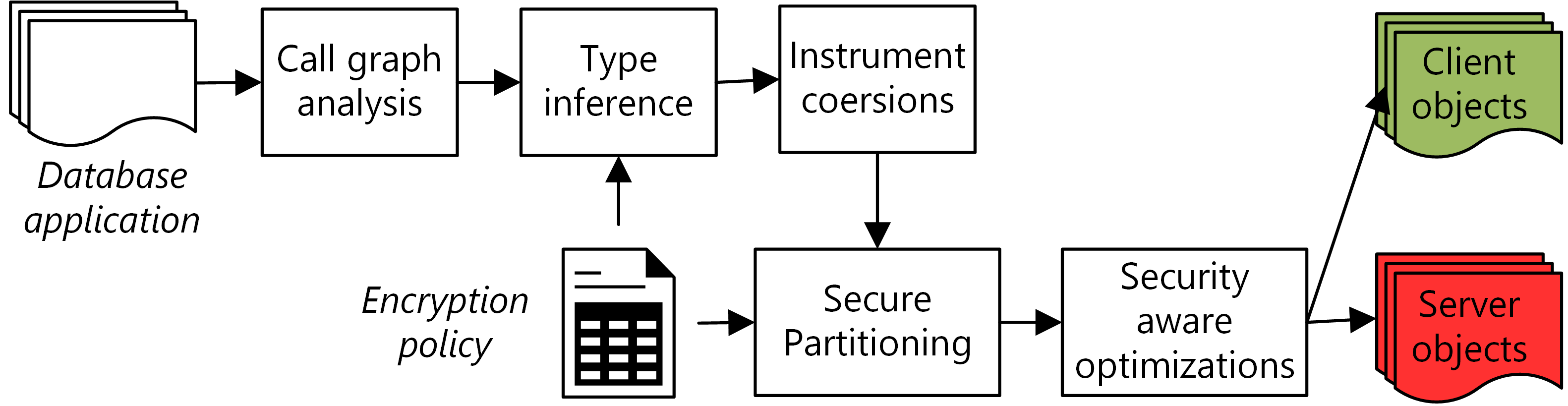}
\caption{
\label{fig:toolchain}
The tool chain for partitioning database applications. 
The input to the tool chain are a database application and an encryption policy.}
\vspace{-0.15in}
\end{figure}

The core component of \securesql\ is a query compiler for T-SQL (Figure~\ref{fig:toolchain}). 
The inputs to the compiler are a database application consisting of a set of queries and stored procedures, and an \textit{encryption policy}, which specifies the set of columns to be encrypted, and the encryption type, encryption algorithm and encryption key for each column.  
The compiler partitions queries and stored procedures into client and server components while preserving semantics of the application {\it and} ensuring the absence of insecure information flows. 
The compiler is based on a type system with knowledge of the computational capabilities of partially homomorphic encryption schemes and tracks information flows. 
The compiler also supports several optimizations for efficiently partitioning queries that are beyond the scope of conventional database optimizers. 

We have evaluated \securesql\ (Section~\ref{sec:evaluation}) using TPC-C, a standard OLTP benchmark, and a real world employee performance management application. 
Our evaluation suggests that \securesql\ can guarantee absence of explicit information flows with reasonable performance overheads ($<$ 30\%). 
However, if absence of implicit flows is desired, or if computation on the critical path cannot be performed on encrypted data, performance can degrade (by as much as two orders of magnitude). 
Factors that contribute to the loss in performance include the inability to use indexes on columns encrypted using semantically secure encryption and additional round trips and data transfers between the client and the server.
Users must therefore choose their encryption policy wisely.

The rest of this paper is organized as follows.
We define the threat model for systems based on a trusted client in Section~\ref{sec:problem}.
We reduce the problem of securely partitioning an application into two sub-problems. 
The first problem is to rewrite the application so that its semantics are preserved when columns in the database are encrypted while ensuring the absence of information flows. 
We describe a type system that achieves this in Section~\ref{sec:types}. 
The second problem is to partition the rewritten application between the client and server while preserving information flow safety (Section~\ref{sec:partitioning}). 
We present an evaluation in Section~\ref{sec:evaluation} and conclude in Section~\ref{sec:conclusions}. 
\section{Threat model}
\label{sec:problem}

We wish to protect sensitive data from an honest-but-curious adversary who has access to contents of the database and all queries. 
The adversary can observe the state of the server on disk, in memory and all communication over the network.
The adversary, however, does not have access to encryption keys, which are stored on the trusted client. 
\securesql\ does not protect against \textit{active} adversaries who can tamper with code and data.
While certain kinds of integrity attacks can be detected using authenticated encryption~\cite{authenticated-encryption}, ensuring end-to-end integrity of query processing is an open problem beyond the scope of this paper. 
We also preclude adversaries who exploit side channels such as size of inputs and results, address traces, timing, and power consumption. 

In \securesql, the adversary can gain information about sensitive columns due to (a) use of weaker encryption schemes which are not semantically secure, and (b) information flows. 
As discussed earlier, weaker encryption schemes are a core component of encryption databases - they allow computation to be offloaded to the server, and make use of database optimizations such as indexes. 
To permit these schemes, we must weaken the threat model, and consider any information revealed due to weak encryption as prior knowledge known to the adversary.
Specifically, for deterministic encryption, we assume that the adversary already knows whether two cipher text values correspond to the same cleartext.
In cryptographic proofs of security, this is achieved by restricting the adversary from encrypting the same value more than once; see ~\cite{bellare07deterministic, amanatidis07new} for formal definitions of security of deterministic encryption. 
Similarly, we assume that the adversary is already knows the relative ordering of values in a column encrypted using order-preserving encryption. 
Our goal is  to ensure that query processing does not reveal any \textit{additional} information. 

\label{sec:overview}
\begin{figure}[!t]	
\lstset{basicstyle=\ttfamily\scriptsize,tabsize=1,numbers=left,language=sql}
\begin{lstlisting}
CREATE PROCEDURE [dbo].[PAYMENT]
@c_w_id INT, @h_amount NUMERIC(6,2), @c_last CHAR(16)
AS BEGIN
	BEGIN TRANSACTION;
		UPDATE dbo.CUSTOMER
		SET @c_id = C_ID, 
			@c_first =  C_FIRST, @c_credit = C_CREDIT,
			@c_balance = C_BALANCE = C_BALANCE + @h_amount,
		WHERE CUSTOMER.C_W_ID = @c_w_id 
		AND CUSTOMER.C_LAST = @c_last;
		
		INSERT dbo.HISTORY (H_C_ID, H_C_BALANCE)
		VALUES (@c_id, @c_balance)		

		IF @c_credit = 'BC'
			UPDATE dbo.CUSTOMER
			SET C_DATA = @TIMESTAMP + C_DATA
			WHERE CUSTOMER.C_W_ID = @c_w_id
			AND CUSTOMER.C_LAST = @c_last;

		SELECT @c_id AS N'@c_id', @c_first AS N'@c_first',
			@c_last AS N'@c_last', @c_credit AS N'@c_credit',
			@c_balance AS N'@c_balance',
	COMMIT TRANSACTION;
END
\end{lstlisting}
\caption{
\label{fig:tpcc-payment-orig}
A T-SQL procedure derived from TPC-C}
\end{figure}

The second source of information leakage is insecure information flows through the trusted client.
We illustrate such leakage using an example. 
Consider the TPC-C benchmark~\cite{tpcc}, an application that maintains customer, inventory, and order processing information.  
Figure~\ref{fig:tpcc-payment-orig} shows a stored procedure derived from TPCC. 
This procedure records payments received from a customer. 
For this stored procedure, we can define an encryption policy as follows. 
Since customer information is sensitive, we should ideally encrypt all personally identifiable information (PII) in the customer table (such as names, address details, account balance and credit rating) using randomized encryption. 
However, the column {\tt CUSTOMER.C\_BALANCE} is involved in an addition, and columns {\tt CUSTOMER.C\_LAST} and {\tt CUSTOMER.C\_CREDIT} are used in equality checks. 
Since randomized encryption prohibits these operations, we can use Paillier encryption (which is semantically secure) for {\tt C\_BALANCE} and deterministic encryption for {\tt C\_LAST} and {\tt C\_CREDIT}. 
Deterministic encryption also allows the query engine to maintain and use an index on these columns. 
We can leave all other non-PII columns (such as inventory and order information) unencrypted. 

While this policy appears to protect all sensitive information, it does not guarantee confidentiality (even under the relaxed threat model discussed above).
\begin{itemize}
\item 
There is an explicit flow from {\tt CUSTOMER.C\_BALANCE} to {\tt HISTORY.C\_BALANCE}, which is not encrypted. 
The flow reveals account balances even if the adversary does not have access to encryption keys.
Preventing this flow requires a simple change to the policy - encrypt {\tt HISTORY.C\_BALANCE} using Paillier encryption. 
\item
There is an implicit flow from {\tt CUSTOMER.C\_CREDIT} to {\tt CUSTOMER.C\_DATA} in the update query at line 17 because {\tt C\_DATA} (which is not encrypted) is conditionally updated based on {\tt C\_CREDIT}. 
Due to this flow, an adversary can potentially infer whether a customer has a bad credit history. 
This flow can be prevented by encrypting {\tt C\_DATA} using deterministic or randomized encryption. 
\end{itemize}

\begin{figure}[!t]	
\lstset{basicstyle=\ttfamily\scriptsize,tabsize=1,numbers=left,language=sql,xleftmargin=5ex}
\begin{lstlisting}
--Server
CREATE PROCEDURE [dbo].[__Closure]
@c_w_id INT, @h_amount VARBINARY(2048), @c_last VARBINARY(256)
AS BEGIN
	SELECT @pubkey = ...
	BEGIN TRANSACTION;
		UPDATE dbo.CUSTOMER
		SET @c_id = C_ID,
			@c_first =  C_FIRST, @c_credit = C_CREDIT,
			@c_balance = C_BALANCE = 
				dbo.PaillierAdd(C_BALANCE, @h_amount, @pubkey)
		WHERE CUSTOMER.C_W_ID = @c_w_id
			AND CUSTOMER.C_LAST = @c_last;

		INSERT dbo.HISTORY (H_C_ID, H_C_BALANCE)
		VALUES (@c_id, @c_balance)		

		IF @c_credit = 0x002057E9A8865AAA7D59DA69AD...
			UPDATE dbo.CUSTOMER
			SET C_DATA = @TIMESTAMP + C_DATA
			WHERE CUSTOMER.C_W_ID = @c_w_id
			AND CUSTOMER.C_LAST = @c_last;

		SELECT @c_id AS N'@c_id', @c_first AS N'@c_first',
			@c_last AS N'@c_last', @c_credit AS N'@c_credit',
			@c_balance AS N'@c_balance'
	COMMIT TRANSACTION;
END

-- Trusted Client
CREATE PROCEDURE [dbo].[PAYMENT]
@c_w_id INT, @h_amount NUMERIC(6,2), @c_last CHAR(16)
AS BEGIN
	SELECT @key = ... // private key 
	SELECT @pubkey = ...	// public key for paillier
	SELECT 
		@enc_h_amount = dbo.AEncrypt(@h_amount, @key, @pubkey),
		@enc_c_last = dbo.DEncrypt(@c_last, @key),
	
	EXEC [SERVER].[tpcc].dbo.__Closure 
		@enc_c_w_id, @c_d_id,@enc_c_amount, @enc_c_last,
		out @c_id, out @c_first, out @c_last, 
		out @c_balance, out @c_credit	
	
	SELECT @c_id AS @c_id,
		dbo.RDecrypt(@c_first, @key) AS @c_first,
		dbo.DDecrypt(@c_last, @key) AS @c_last,
		dbo.DDecrypt(@c_credit, @key) AS @c_credit,
		dbo.ADecrypt(@c_balance, @key, @pubkey) AS @c_balance,
END
\end{lstlisting}
\caption{
\label{fig:tpcc-payment-rewritten}
A partitioned stored procedure which preserves semantics and absence of explicit flows.}
\end{figure}

Clearly, these flows are not desirable. 
We now define a security property that disallows insecure information flows. 
\begin{definition}
\label{def:secure}
A application preserves confidentiality if there are no information flows from a column/variable encrypted using a \textit{stronger} encryption scheme to a column/variable encrypted using a \textit{weaker} encryption scheme. 
\end{definition}

Note that this definition assumes the presence of a partial order over encryption schemes based on their relative strength. 
Also note that this definition includes all columns and variables such as parameters passed to any procedures executed on the server, and results of intermediate query processing. 
We can extend this definition to an application partitioned into a trusted client and an untrusted server. 
\begin{definition}
\label{def:secure-partitioning}
A partitioned application preserves confidentiality if there are no information flows from a column/variable encrypted using a stronger encryption scheme to a column/variable \textit{stored on the server} encrypted using a weaker encryption scheme. 
\end{definition}

Unlike Definition~\ref{def:secure}, this property permits information flows from encrypted columns to cleartext variables on the trusted client. 
This allows the trusted client to decrypt values retrieved from encrypted column, perform computations on those values, encrypt the results, and store them on the server \textit{as long as resulting values are encrypted using a sufficiently strong encryption scheme}. 


The example described above also illustrates the security-performance trade-offs that arise in the trusted client model. 
Ensuring the absence of explicit and implicit flows often requires more columns to be encrypted, and additional round trips to the client. 
\securesql's type system identifies such flows, and gives users the choice of enforcing absence of explicit and/or implicit flows.
Figure~\ref{fig:tpcc-payment-rewritten} shows a partitioning generated by \securesql\ for this example when implicit flows are permitted.
The partitioning is efficient since all computation is offloaded to the server. 
The trusted client simply encrypts parameters, uses linked server at line 39 to invoke the server component, and decrypts results. 
Refer to ~\cite{securesql15supplement} for the partitioning that contains no explicit or implicit flows. 
This partitioning requires the column {\tt C\_DATA} to be encrypted and delegates the string concatenation operation in line 18 to the client (at the cost of an additional round trip).
\section{Information flow types for partially homomorphic encryptions}
\label{sec:types}

The type system in \securesql\ serves dual purposes. 
First, it automatically rewrites a database application by inserting calls to encryption/decryption routines so that its semantics are preserved in the database with encrypted columns. 
Furthermore, the type system ensures the absence of explicit and/or implicit flows, assuming all local variables are stored in trusted memory. 
The partitioning algorithm (Section~\ref{sec:partitioning}) further partitions stored procedures, moving local variables and statements from the trusted client to the server while preserving information flow safety. 
While type systems for information flow have been studied extensively~\cite{dennis97typebasedsecurity, sabelfeld03language-basedinformation-flow}, the combination of partially homomorphic encryption and information flow, and its use for automatic partitioning are unique to our system.\\

\begin{figure}[t]
\[
\begin{array}{l}
{
\begin{array}{lllllllll}
\name{} & \in & \textsf{Names} & \val & \in & \textsf{Values} & \var & \in & \textsf{Variables}\\
\func & \in & \textsf{Functions} & \tabl{} & \in & \textsf{Tables} & \column{} & \in & \textsf{Columns} \\
\key & \in & \textsf{Keys} & pk & \in & \textsf{PublicKeys} & S & \in & \textsf{Server}\\
\end{array}
}
\\
\\
{
\begin{array}{rll}
\expr & ::= \val \mid \var \mid \tablexpr{}.\column{} \mid \namedrec \mid \exprsub{1} = \exprsub{2} \mid \exprsub{1} < \exprsub{2}\\
& \mid \exprsub{1} + \exprsub{2} \mid \func\expr \mid \selectexpr \mid \projectexpr \mid \unionexpr \mid \diffexpr \mid \productexpr\\
\\
\stmt & ::= \exprsub{1} := \exprsub{2}  \mid \stmt_{1}; \stmt_{2} \mid \ifexpr \mid \selectstmt\\
& \mid \insertexpr \mid \updateexpr \mid \deleteexpr\\
\\
\prog & ::= \func(\var).\stmt \mid \prog\ \prog &\\
\\
\expr_t & ::= \ndencrypt(\key,\var) \mid \nddecrypt(\key,\var) \mid \dencrypt(\key,\var)\\
& \mid \ddecrypt(\key,\var) \mid \opencrypt(\key,\var) \mid \opdecrypt(\key,\var)\\
& \mid \aencrypt(\key, pk,\var) \mid \adecrypt(\key,pk,\var) \mid [S].\stmt
\end{array}
}
\end{array}
\]
\caption{
\label{fig:lsql-syntax}
Syntax of \LSQL\ 
}
\end{figure}

\paragraph{Language.}
Instead of presenting the type system for the entire T-SQL language, we define a simpler, core language called \LSQL\ (Figure~\ref{fig:lsql-syntax}) which models key features of declarative query languages. 
An \LSQL\ program \prog\ is a collection of named stored procedures.
The state of a \LSQL\ program consists of a database with a set of tables $t$ and a set of local variables $v$. 
The body of a procedure is a statement ($s$).
An expression is either a constant (\val), a variable ($v$), reference to a column in a table (\tablexpr{}.\column{}) or a named record \namedrec. 
A named record is a tuple with a unique name associated with each element. 
Named records permit relational operations such as projection and cartesian product. 
\LSQL\ supports basic arithmetic, boolean operations and function application, assignment,  imperative control flow in the form of branching statement, standard relational operators for querying data including projection (\projectexpr), selection (\selectexpr), union (\unionexpr), difference (\diffexpr), and cartesian product (\productexpr). 
\projectexpr\ selects a subset of columns from a collection of named records. 
\selectexpr\ selects all tuples from the result of the expression \exprsub{2} that satisfy the predicate \exprsub{1}. 
Tables ($t$) can be modified using \ins, \update, and \delete\ statements.
The \select\ statement can be used to query tables and return results.

\LSQL\ also supports a set of expressions used in generated code but not available to the programmer ($\expr_t$). 
This includes the implementations of encryption and decryption algorithms. 
The algorithms for randomized, deterministic and order-preserving encryption use symmetric private keys, whereas additive encryption use asymmetric, public-private key pairs. 
We assume that the algorithm for randomized and additive encryption are secure under standard cryptographic assumptions. 
The language does not support explicit creation or manipulation of encryption keys - we assume that the application (i.e. the trusted client) has access to a finite number of predefined keys/key pairs. 
Finally, $[S].\stmt$ offloads execution of statement $\stmt$ from a client database to a server $S$.\\

\begin{figure}[!t]
\[
\begin{array}{rll}
\typevar{} & ::= & \spt \mid \snde \mid \sade \mid \sde \mid \sope \mid \void\\
&& \mid \rectype \mid \namedrectype \mid \typevar{1} \rightarrow \typevar{2}
\end{array}
\]
\caption{
\label{fig:type-system}
Types in the type system for partially homomorphic encryptions.}
\end{figure}

\paragraph{Types.}
Given an \LSQL\ program, our goal is to assign each expression in the program a type which represents whether the value returned by the expression is encrypted or not, and the type of encryption used i.e. randomized (\snde), additive (\sade), deterministic (\sde) or order-preserving (\sope).
\securesql's type system (Figure~\ref{fig:type-system}) consists of a set of primitive types, one for each type of encryption. 
The type \spt\ represents cleartext values, and the type $\void$ represents expressions that do not return values (e.g. assignment). 
The type system also supports function types and named record types. 
A named record type is a type representing a tuple of values, where each element in a record type is associated with a name.
Record types are a natural way of representing types of tables and relational operators (such as projection and product). 

Note that types described above do not refer to encryption keys. 
For the purpose of this paper, we make a simplifying assumption that all values encrypted using the same kind of encryption use the same encryption key. 
In our implementation, a type consists of the pair $<$encryption type, keyid$>$, where keyid is a unique identifier representing each encryption key.\\

\begin{figure}[t]
\centering
\subfloat{}{}{\includegraphics[width=1.3in]{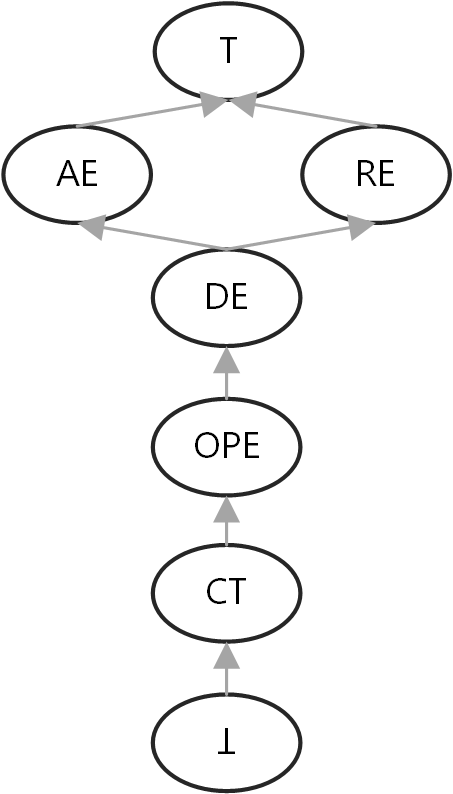}}
\subfloat{}{}{\includegraphics[width=2in]{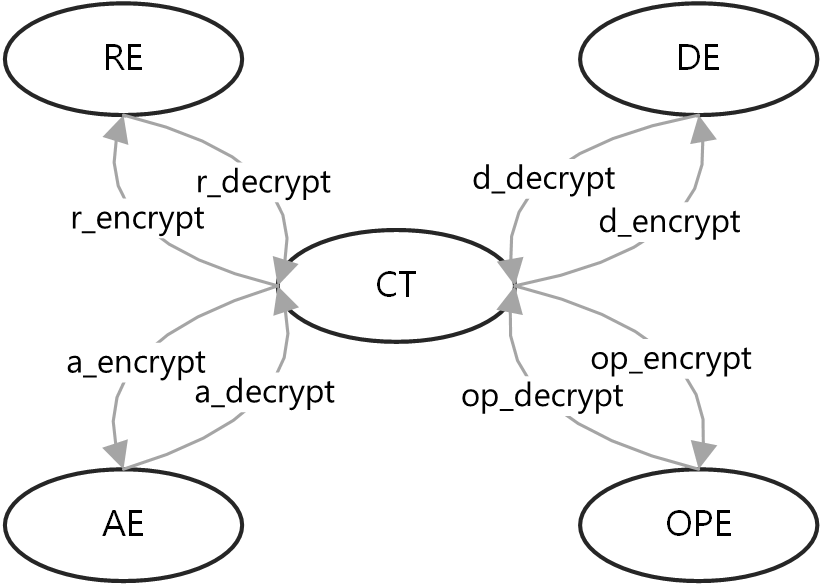}}
\caption{
\label{fig:subtyping-relation}
Types and encryption and decryption routines for coercing values between types. 
}
\vspace{-0.1in}
\end{figure}

\paragraph{Encryption/decryption as Coercions.}
In a conventional type system for information flows~\cite{dennis97typebasedsecurity, sabelfeld03language-basedinformation-flow}, types correspond to secrecy levels in a lattice. 
For example, the types $\top$ and $\bot$ represent secret and public values respectively. 
In an encrypted database, secrecy levels correspond to partially homomorphic encryption schemes. The encryption schemes can be ordered by the relative strength of encryption; this ordering forms a lattice shown in Figure~\ref{fig:subtyping-relation}. 
A type system based on this lattice can be used to \textit{check} if a partitioned application contains any explicit or implicit information flows, 
and if the application uses partially homomorphic encryptions correctly. 
However, just type checking does not suffice if our starting point is a monolithic application written for a database without encryption -- we wish to instrument encryption/decryption routines to preserve semantics of the application, and partition the application while preventing information flows. 

Towards this end, we define a type system where encryption/decryption routines are viewed as \textit{coercions} between types, similar to how compilers coerce of values from one (usually less precise) type to another (more precise) type. 
Specifically, we propose a type system with coercive subtyping~\cite{luo99coercive}. 
In coercive subtyping, \typevar{1} is said to be a subtype of \typevar{2} if a value of type $\typevar{1}$ can be coerced into a value of type $\typevar{2}$ using a coercion function $\coersionfunc_{\typevar{1}\rightarrow\typevar{2}}$. 
In such a system, the {\it subtyping relation} \subtyperel\ defines the set of coercions permitted by the type system. 
In our type system, encryption and decryption routines are coercion functions.
Figure~\ref{fig:subtyping-relation} shows all coercion functions along with the source and target types.
Essentially, we can coerce a value of a given encryption type to a value of any other encryption type using one or more encryption/decryption routines. 

At first glance, the relation in Figure~\ref{fig:subtyping-relation} appears to be an obvious choice for a subtyping relation. 
However, permitting coercions between all encryption types is not desirable.
If we permit arbitrary coercions, values from encrypted columns can be coerced into values encrypted using weaker encryption schemes, resulting in insecure information flows. 
Another reason for not permitting all coercions is efficiency -- type inference is tractable only when the subtyping relation is a partial order, and efficient when the partial order is a lattice~\cite{pratt82satisfiability}.

\begin{figure}[t]
\centering
\includegraphics[width=3in]{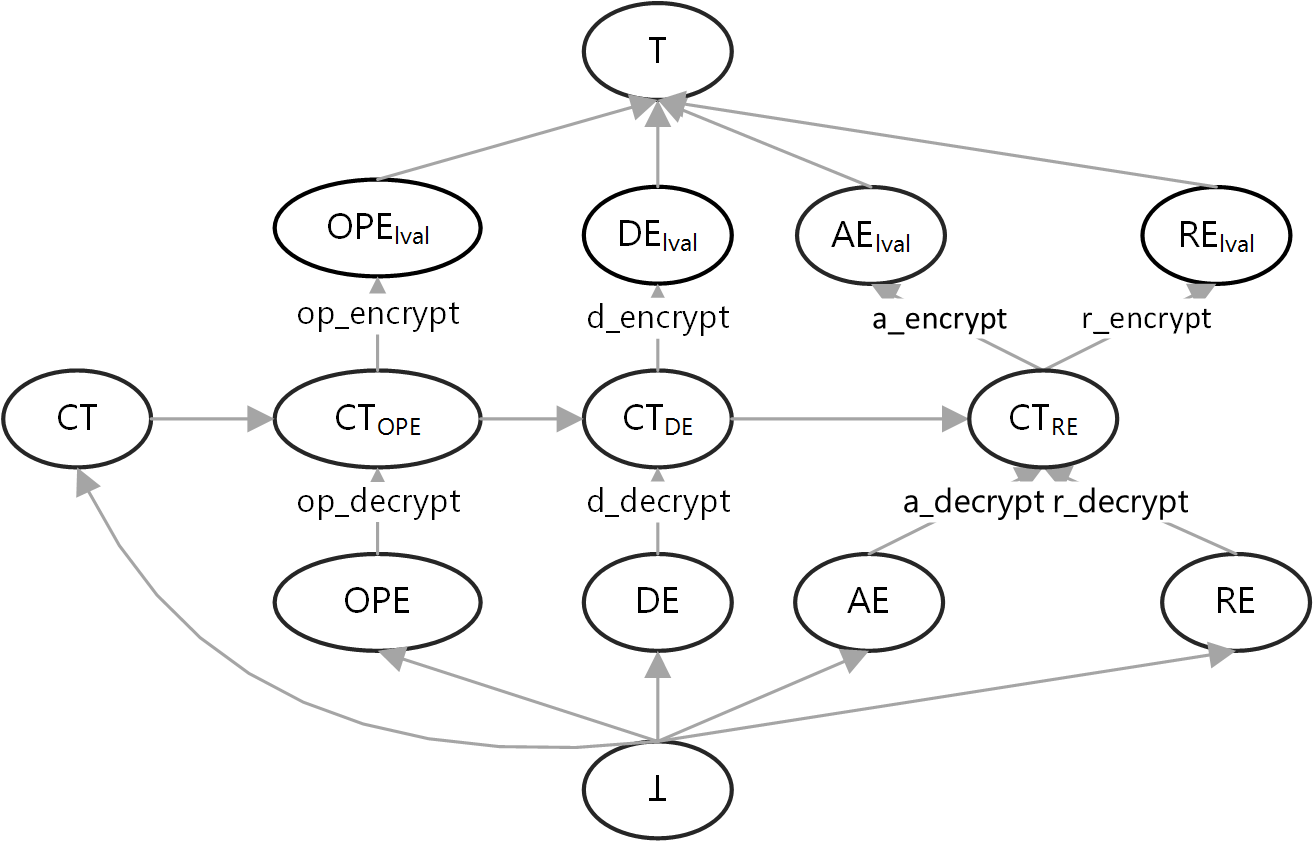} 
\caption{
\label{fig:unrolled-subtyping-relation}
A coercive subtyping relation which tracks information flow from encrypted types.
Unlabeled edges have identity functions as coercions.
}
\vspace{-0.1in}
\end{figure}

Figure~\ref{fig:unrolled-subtyping-relation} shows the subtyping relation used in \securesql.
This relation (which is a lattice) is derived from Figure~\ref{fig:subtyping-relation} by unrolling cycles once and adding a few additional types. 
The types \oppt, \dpt, and \ndpt are variants of the type \spt\ -- they distinguish cleartext values obtained via decryption from cleartext values that were never encrypted. 
Both randomized and additive encryption share the same clear text type because they are both equally secure (in the cryptographic sense). 
$\sopel, \sdel, \sadel$, and $\sndel$ are variants of encrypted types which represent re-encrypted values. 
Unlabeled edges have identity functions as coercions.
Observe that this subtyping relation permits encrypted values to be decrypted into cleartext values, and encrypted back using the same or stronger encryption scheme. 
As described in Section~\ref{sec:partitioning}, these additional types allow the partitioning algorithm to identify computations that can be safely offloaded to the server. \\



\paragraph{Type inference.}
We now describe \securesql's algorithm (based on ~\cite{mitchell91typeinference}) for automatically inferring encryption types and coercions. 
Informally, the algorithm work as follows.
The algorithm associates two \textit{type variables} with every expression, an \textit{assumed type} $\alpha$, which represents the type of the expression without coercions, and an \textit{expected type} $\beta$, which represents the type after a coercion permitted by the subtyping relation $\acylsubtyperel$ has been applied. 
The algorithm collects set constraints on these type variables during a traversal of the AST.
A solution to these constraints, if one exists, assigns values to assumed and expected type variables of every expression. 
Expressions which require requires coercions are expressions with different assumed and expected types.

For tracking implicit flows, the algorithm associates a type variable $\gamma$ with every statement $s$.
$s$ has an encryption type $\gamma$ if all columns/variables updated by $s$ have encryption type at least as strong as $\gamma$. 
$\gamma$ also represents the {\it context} in which a statement can execute without introducing implicit flows. 
A statement $s$ with type $\gamma$ can safely execute in a context of type $\gamma$ or weaker. 

The algorithm can be stated as a set of type rules. 
Figure~\ref{fig:type-inference-sample} shows a subset of the type rules (refer to ~\cite{securesql15supplement} for the complete set of type rules). 
Type judgments for expressions take the form $\constraints,\typeenv\satisfies\expr:\beta$. $\constraints$ is a set of type constraints of the form $\tau \subtyperel \beta$ or $\tau \in S$, where $\tau$ is a type expression, $\beta$ is a type variable, and $S$ is a set of types. 
$\typeenv$ is a map from expressions to their assumed types.
A type judgment is interpreted as follows: under constraints $\constraints$ and assumptions $\typeenv$, the expected type of $\expr$ is $\beta$. 
Type judgments for statements take the form $\constraints,\typeenv,\gamma\satisfies\stmt$, where $\gamma$ represents the context in which the statement $\stmt$ can safely execute. 

Consider the rule VAR for variables.
The rule allocates two fresh type variables $\alpha$ and $\beta$, sets the assumed type of $\var$ to $\alpha$, and adds a constraint that $\alpha\subtyperel\beta$, where $\beta$ is the expected type of $\var$. 
The rule for column references is similar, with the only difference that the assumed type of a column reference is obtained from the encryption policy \policy. 

Now consider the rule EQUALS.
Let $\beta_1$ and $\beta_2$ be the expected types of $\exprsub{1}$ and $\exprsub{2}$ respectively. 
Equality checking requires types of both sub-expressions to be the same.
This is enforced by unifying both types (using the function \unify). 
Equality also requires that both types should be at least at weak as deterministic encryption. 
We enforce this using a subset constraint $\beta_{1}\in [\sope, \sopel, \sde, \sdel, $ $\spt, \oppt, \dpt, \ndpt]$. 
The algorithm also unifies the assumed types of variables common to $\exprsub{1}$ and $\exprsub{2}$. 
The constraint on the type of equality check reflects the fact that unlike other partially homomorphic encryptions, deterministic encryption and OPE reveal the result of the check in cleartext. 
The rule COMP for comparison follow the same template and only differ in the set of permitted encryption types. 

Inserting into a table (INSERT) requires that the type of values being inserted must be a subtype of the column being written to (as defined in the policy).  
Here, we use the function \lvaltype\ to obtain the re-encrypted type corresponding to an encryption type. 
Observe that this rule allows re-encryted values to be written to a column to a column of a given encryption type.
For example, a value of type \sndel\ obtained by decrypting a value of type \sdel\ and re-encrypting it using randomized encryption can be written to a column of type \snde. 
The type $\gamma$ captures the strongest encryption type that is written by the statement. 

\begin{figure}[!t]
\small
\begin{mathpar}
\inferrule* [Lab={\tiny [CONST]}]
{\beta=\fresh}
{\{\spt \subtyperel \beta\},\{\val : \spt\}\satisfies \val : \beta}
\and
\inferrule* [Lab={\tiny [VAR]}]
{\alpha,\beta=\fresh}
{\{\alpha \subtyperel \beta\},\{\var : \alpha\}\satisfies \var : \beta}
\and
\inferrule* [Lab={\tiny [COLUMN]}]
{\alpha=\policy(\tablexpr{}, \column{})\judgespace\beta=\fresh}
{\{\alpha \subtyperel \beta\},\{\var : \alpha\}\satisfies \tablexpr{}.\column\ : \beta}
\and
\inferrule* [Lab={\tiny [EQUALS]}]
{\judgeexpr{1}\judgespace\judgeexpr{2}\judgespace\beta=\fresh \\\\ \subs=\unify(\{\alpha,\beta \mid \var : \alpha \in \typeenv_{1} \wedge \var : \beta \in \typeenv_{2}\} \cup \{\beta_1, \beta_2\})}
{\subs(\constraints_{1})\cup\subs(\constraints_{2}) \cup \{\beta\in [\spt,\oppt,\dpt,\ndpt]\} \cup\\\\\{\beta_{1}\in [\sope,\sopel,\sde,\sdel,\spt,\oppt,\dpt,\ndpt],\\\\\subs({\beta_1})\subtyperel\beta\},\subs(\typeenv_{1})\cup\subs(\typeenv_{2})\satisfies\exprsub{1} = \exprsub{2} : \beta}
\and
\inferrule* [Lab={\tiny [COMP]}]
{\judgeexpr{1}\judgespace\judgeexpr{2}\judgespace\beta=\fresh \\\\ \subs=\unify(\{\alpha,\beta \mid \var : \alpha \in \typeenv_{1} \wedge \var : \beta \in \typeenv_{2}\} \cup \{\beta_1, \beta_2\})}
{\subs(\constraints_{1})\cup\subs(\constraints_{2}) \cup \{\beta\in [\spt,\oppt,\dpt,\ndpt]\} \cup\\\\\{\beta_{1}\in [\sope,\sopel,\spt,\oppt,\dpt,\ndpt],\\\\\subs({\beta_1})\subtyperel\beta\},\subs(\typeenv_{1})\cup\subs(\typeenv_{2})\satisfies\exprsub{1} < \exprsub{2} : \beta}
\and
\inferrule* [Lab={\tiny [ADD]}]
{\judgeexpr{1}\judgespace\judgeexpr{2}\judgespace\beta=\fresh\\\\\subs=\unify(\{\alpha,\beta \mid \var : \alpha \in \typeenv_{1} \wedge \var : \beta \in \typeenv_{2}\} \cup \{\beta_1, \beta_2\})}
{\subs(\constraints_{1})\cup\subs(\constraints_{2})\cup\{\beta_1 \in [\sade,\sadel,\spt,\oppt,\dpt,\ndpt],\\\\\subs({\beta_1})\subtyperel\beta\},\subs(\typeenv_{1})\cup\subs(\typeenv_{2})\satisfies\exprsub{1} + \exprsub{2} : \beta}
\and
\inferrule* [Lab={\tiny [ASSIGN]}]
{\constraints_{1},\typeenv_{1}\cup \{\exprsub{1} : \alpha_{1}\}\satisfies\exprsub{1} : \beta_{1}\judgespace\judgeexpr{2}\\\\\subs=\unify(\{\alpha,\beta \mid \var : \alpha \in \typeenv_{1} \wedge \var : \beta \in \typeenv_{2}\}\cup\{\alpha_{1},\beta_{1}\})}
{\subs(\constraints_{1})\cup\subs(\constraints_{2})\cup\{\subs(\beta_{2})\subtyperel\subs(\beta_{1})\},\subs(\typeenv_{1})\cup\subs(\typeenv_{2}),\subs(\beta_{1})\satisfies\exprsub{1}:=\exprsub{2}}
\and
\inferrule* [Lab={\tiny [INSERT]}]
{\forall{i},\ \beta_{i}=\lvaltype({\policy(\tablexpr{},\column{i})})\judgespace\judgeexpr{}\judgespace\gamma=\fresh}
{\constraints\cup\{\beta\subtyperel[\beta_1,\dots,\beta_n]\cup\{\forall{i}\in[1,\dots,n],\subs(\beta_{i})\subtyperel\gamma\},\typeenv,\gamma\satisfies: \insertexpr}
\and
\inferrule* [Lab={\tiny [SELECT]}]
{\judgeexpr{}\judgespace\forall{i}\in[1,\dots,m],\ \beta_{i}=\project(\beta, \name{i})\judgespace\gamma=\fresh}
{\{\constraints\cup\{\forall{i}\in[1,\dots,n],\subs(\beta_{i})\subtyperel\gamma\},\typeenv,\gamma\satisfies: \selectstmt}
\and
\inferrule* [Lab={\tiny [IF]}]
{\judgeexpr{1}\judgespace\judgestmt{1}\judgespace\judgestmt{2}\\\\\subs=\unify(\{\alpha,\beta \mid \var : \alpha \in \typeenv_{i} \wedge \var : \beta \in \typeenv_{j}\} \cup \{\gamma_{1}, \gamma_{2}\})}
{\subs(\constraints_{1})\cup\subs(\constraints_{2})\cup\subs(\constraints_{3})\cup\{\subs(\beta_{1})\subtyperel\subs(\gamma_{1})\},\\\\\subs(\typeenv_{1})\cup\subs(\typeenv_{2})\cup\subs(\typeenv_{3}),\subs(\gamma_{1})\satisfies\ifexpr}
\end{mathpar}
\caption{
\label{fig:type-inference-sample}
A subset of rules from the type inference algorithm.}
\vspace{-0.1in}
\end{figure}

Consider the rule IF.
The constraint on the contexts of 'then' and 'else' statements these statements can only update columns encrypted with a stronger scheme than the variables involved in the branching condition. 
For example, consider the IF statement at line 16 in Figure~\ref{fig:tpcc-payment-orig}. 
If the column {\tt C\_CREDIT} is encrypted using deterministic encryption, the equality check can be performed on the server by encrypting the constant {\tt 'BC'}. 
In this case, the type of the branching condition is \dpt\ (rule EQUALS).
The rule IF ensures that all columns updated in the 'then' branch should have a type at least as strong at \dpt. 
If the column {\tt C\_DATA} is not encrypted, this condition cannot be satisfied due to an implicit flow, and the algorithm reports an error.


\paragraph{Parameters and return values.}
The algorithm introduces additional constraints on parameters and return values of procedures that can be externally invoked. 
Specifically, we require that the type of all parameters and the return variable of such procedures must be one of the cleartext types. 
This constraint ensures that the interface of these stored procedures does not change from the perspective of external callers.
\\

\paragraph{Procedure and function calls.}
The \securesql\ compiler handles procedure calls as follows.
Given a set of stored procedures, the compiler constructs a call-graph, performs a bottom up traversal of the call-graph and runs the type inference algorithm described above for each procedure. 
After processing each node, computes a \textit{summary} for each procedure, which consists of the encryption types of the formal parameters and return values.
The summary also records whether the stored procedure requires any coercions. 
At a procedure call, the compiler uses the summary to introduce constraints on parameters and return values. 
Specifically, we require that the expected type of an actual parameter should be a subtype of the corresponding formal parameter, and the type of the return value should be a subtype of the expected type of the actual return value. 
In case of recursive calls, the compiler repeats the inference until a fixed point is reached.\\



\paragraph{Cost-aware constraint solving.}
The constraints generated by the type system are inequalities over the subtyping relation, which is a partial order. 
In general, the problem of poset solving is NP-complete~\cite{pratt82satisfiability}. 
However, our subtyping relation is a lattice with the join defined by the lowest common ancestor relation. 
Solving inequalities over a lattice is linear in the number of expressions and the height of the lattice~\cite{pratt82satisfiability}.
The result of constraint solving is either a type assignment which assigns a type to each type variable, or a set of unsatisfiable constraints. 
A set of constraints may be unsatisfiable if the program contains insecure information flows. 
We report all unsatisfiable constraints to the user, and also recommend the changes to the encryption policy that will satisfy the constraints.  

One shortcoming of the inference algorithm described above is that it unaware of the runtime cost of coercions. 
For example, consider the equality check at line 10 in Figure~\ref{fig:tpcc-payment-orig}. 
If the column {\tt CUSTOMER.C\_LAST} is deterministically encrypted, there are two possible assignments of types to expressions {\tt CUSTOMER.C\_LAST} and {\tt @c\_last} that will preserve semantics of the filter, one in which both expressions are assigned the type \dpt, and the other where they are assigned the type \sdel. 
Both type assignments result in the same number of coercions (i.e. one). 
In the first case, the column {\tt C\_LAST} is decrypted, and in the second case, the variable {\tt @c\_last} is deterministically encrypted. 
The resulting queries have very different query plans. 
In the first case, the column {\tt C\_LAST} must transferred to the client and the check is performed on the client. 
The second case lends to a more efficient plan where the filter is performed on the server using an index.
  
In order to find efficient type assignments, we extend the inference algorithm in ~\cite{pratt82satisfiability} with a simple cost model and a procedure that systematically explores the space of all valid type assignments to find type assignments with minimal cost.  
Our cost model computes the cost of a type assignment by assigning a cost to all expressions that must be coerced under the assignment. 
The cost of a coercion on a column reference is approximated by the cardinality of the column (estimated using statistics from the server). The cost of coercing as primitive value is 1. 
The cost of an expression is the sum of the cost of its sub-expressions. 

We explore the space of all valid type assignments using the following iterative algorithm. 
We start with a valid type assignment obtained using the inference algorithm described above, and systematically derive other type assignments by substituting the type assigned to a type variable with its super-types. 
We use each derived type assignment to assign the \textit{assumed} type of each type variable and rerun the inference algorithm. 
If the inference algorithm succeeds, we have found another valid but different type assignment and we can compute its cost. 
If the inference fails, we ignore the type assignment and continue the search. 
The complexity of the exploration is polynomial in the number of expressions and the height of the subtyping relation. 
\\

\paragraph{Inserting Coercions.}
Given a type assignment, the expressions that require coercions are expressions which generated constraints of the form $\alpha\subtyperel\beta$ where $\alpha\neq\beta$.
Such expressions require a coercion from $\alpha$ to $\beta$ because the assumed and expected type are different. 
For every such constraint, we derive the coercion function by traversing the path from $\alpha$ to $\beta$ in the subtyping relation and composing coercions functions for each edge on the path, and instrument the expression with the coercion function.
We also identify all addition expressions where both sub-expressions are assigned a type \sade\ or \sadel, and replace each such addition with a call to homomorphic addition routine. \\

\section{Secure Partitioning}
\label{sec:partitioning}
The next phase in the compilation process is to generate a secure and efficient partitioning of a database application that offloads as much computation as possible to the server without leaking sensitive information. 
In this section, we describe the analysis and optimizations that together generate such partitionings. 
\\

\paragraph{Baseline partitioning.}
The first step in the algorithm is to generate a \textit{baseline partitioning} which preserves semantics. 
For the baseline partitioning,\securesql\ relies on named references, which is a commonly abstraction for distributed query processing in most databases. 
For example, SQL Server supports a primitive called \textit{linked server}~\cite{linkedserver13} and Oracle supports a similar primitive called database link~\cite{databaselink}. 
A linked server is a named reference that can be used by a client database to issue queries on remote servers. 
For example, a client database can issue a query {\tt SELECT * FROM [LINK].[DB].CUSTOMER ORDER BY C\_BALANCE} to select customers from a remote table\footnote{Here {\tt LINK} is the name of the linked server.}.
The query optimizer on the client partitions queries that use linked server between remote servers and the client, utilizing any statistics that may be available on the remote server. 

\securesql\ uses linked server to partition a stored procedure by rewriting all table references to remote table references, and installing the resulting stored procedure on the client. 
Linked server automatically promotes transactions initiated on the client to distributed transactions involving the remote server. 

Unfortunately, while the baseline partitioning is simple and preserves semantics, it is neither secure not efficient. 
In this partitioning, the number of round-trips between the client and server is proportional to the number of queries because the client drives execution, and control returns back to the client after each query, even if the next query can execute entirely on the server. 
Furthermore, transactions are promoted to distributed transactions even if all queries within the transaction are safe. 
The partitioning is also not secure because information can leak through local variables that hold sensitive data in cleartext, and via intermediate results of query processing. 
For example, consider the join query {\tt SELECT * FROM [LINK].R, [LINK].S WHERE dbo.DECRYPT(R.a) = S.a}.
If $\card{R}\ll\card{S}$, a possible query plan is to decrypt the column {\tt R.a} on the client, and transfer the semi-join to the server to compute the join. 
This plan leaks contents of the column {\tt R.a} in plaintext. 
\securesql\ uses a series of analysis and optimizations for generating a secure and efficient partitioning from the baseline. \\

\newcommand{\safe}{{\it safe}}
\newcommand{\safevar}{\ensuremath{\safe_{n}}}
\paragraph{Safety analysis.}
Safety analysis is a static analysis that infers whether a type-annotated T-SQL expression can be safely offloaded to the server without introducing insecure information flows. 
The analysis is based on the following definition for safety. 
A T-SQL expression is \textit{safe} if (a) does not invoke encryption or decryption routines, (b) it does not use any local variables that contain values obtained from prior calls to decryption routines. 

Checking whether an expression invokes encryption/decryption routines is straightforward -- we simply check if any sub-expressions require coercions (using the type assignment). 
We can also check the second condition recursively as follows.
A local variable is {it not} safe if its expected type is one of the types $[\oppt, \dpt, \ndpt]$. 
An expression is safe if all constants and variables in the expression are safe. 
\\

\paragraph{Secure partitioning.}
\securesql\ uses safety analysis to transform the baseline partitioning into a secure (but inefficient) partitioning as follows. 
The compiler identifies and instruments all unsafe scalar expressions with special "identity" routines installed on the client. 
The presence of these routines forces the query optimizer to consider plans where these functions are evaluated on the client. 
For example, consider the query {\tt SELECT * FROM [LINK].R WHERE dbo.IDENTITY(dbo.DECRYPT(R.a) = @filter)}.
Assume that the variable {\tt @filter} is unsafe. 
The call to the identity routine forces the filter to be evaluated on the client. \\


\paragraph{Security-aware optimizations.}
We now describe three optimizations in \securesql, which increase the number of safe expressions, and hence offload more computation to the server.
\securesql\ supports other optimizations such as splitting sub-queries and splitting materialized view definitions -- we omit these due to space constraints. 
A key feature of these optimizations is that they can be expressed as source-to-source transformations, as opposed to transformations on a physical query plan.\\

\noindent\textit{Invariant code motion.}
An expression is \textit{invariant} with respect to a query if its value does not change during the query's execution. 
Invariant code motion is an optimization that identifies invariant expressions (such as calls to encryption/decryption routines), and moves them before the query.
For example, consider the {\tt SELECT} query at line 6 in Figure~\ref{fig:tpcc-payment-orig}.
In this example, the type system infers that variables {\tt @c\_last} and {\tt @h\_amount} should be encrypted, and introduces encryption routines.
However, the expressions {\tt \dencrypt(@c\_last,...)} and {\tt \aencrypt(@h\_amount,...)} are invariant with respect to the query.
The transformation moves these expressions before the query, introduces temporary variables ({\tt @enc\_c\_last} and {\tt @enc\_h\_amount}) to capture the values of these expression, and replaces the expressions with temporary variables. 
The resulting query is free of calls to encryption and decryption routines, and therefore safe.\\
%
%

\noindent
\textit{Offloading Safe Blocks.}
This optimization offloads safe statements to the server. 
The optimization consists of two phases. 
The first phase identifies \textit{maximal safe blocks}. 
A \textit{block} is a set of statements in the same static scope. 
A \textit{maximal safe} block is the largest block such that all statements within the block are safe but at least one of the siblings is not safe. 

The second phase of the transformation partitions the stored procedure further by \textit{extracting} maximal safe blocks into separate stored procedures. 
This is akin to creating closures~\cite{closure}. 
Each maximal safe block is transformed into a stored procedure whose input parameters are variables used within the block, and output parameters are variables defined within the block but used outside.
This transformation then deploys closures on the server, and replaces the maximal safe block with a remote call to the closure.
As a special case, if all statements in a stored procedure are safe, the optimization deploys the entire stored procedure on the server.
Figure~\ref{fig:tpcc-payment-rewritten} shows an instance of this optimization, where the entire procedure is replaced by a call to a closure.\\ 

\noindent
\textit{Distributed transaction elimination.}
Recall that the client database automatically promotes local transactions to distributed transactions while processing queries with remote references.
Distributed transactions are often implemented using expensive, blocking protocols such as 2PC.
This optimization eliminates distributed transactions wherever possible. 
In particular, the optimization checks if all remote accesses within the scope of a transaction occur in one maximally safe block. 
If this condition is satisfied, this optimization eliminate the distribution transaction by reducing the scope of the transaction to the maximal safe block.
This effectively pushes the transaction into the closure, which executes locally on the server. 
Figure~\ref{fig:tpcc-payment-rewritten} shows an example of this transformation.
\newcommand{\performance}{Perf}
\newcommand{\pthop}{PTC}
\newcommand{\pthoptrans}{PTDT}
\newcommand{\posrv}{\textit{NoFlowsSrv}}
\newcommand{\ptsrv}{\textit{NoExpSrv}}
\newcommand{\pthsrv}{\textit{NoExpAddSrv}}
\newcommand{\strongp}{\textit{NoFlows}}
\newcommand{\deterministicp}{\textit{NoExp}}
\newcommand{\phep}{\textit{NoExpAdd}}
\newcommand{\cleartextp}{\textit{ClearText}}

\section{Evaluation}
\label{sec:evaluation}
The goal of our evaluation to is determine the cost of supporting information flow security in encrypted databases. \\

\paragraph{Implementation.}
\label{sec:implementation}
\securesql\ is implemented as a C\# library (\~{} 11000 LOC) based on a publicly available T-SQL parser and code generator~\cite{tsqlparser}.
The type system, rewriting for adding coercions, query partitioning and optimizations are implemented as transformations over the AST generated by the parser. 
We use 256-bit AES block cipher in CBC mode for randomized encryption, AES in GCM mode with an initialization vector derived from the hash of the cleartext for deterministic encryption, 1024 bit Paillier encryption and an order-preserving encryption scheme based on ~\cite{popa2011cryptdb}. \\

\paragraph{Methodology and Benchmarks.}
The experiments were conducted on an in-house cluster connected via a 1 Gbps network. 
Each machine has a 2 Intel Xeon x64 processors with 8 cores each, 16GB RAM, a 1TB SATA drive and runs Windows Server 2008 Enterprise.
Our evaluation is based on two applications TPC-C, and \performance, an in-house employee performance evaluation application. 
The TPC-C database maintains customer, order and inventory information.
We instantiated this database with 100 warehouses (approximately 10 GB).
The workload consists of five transactions implemented as stored procedures. 
The workload generator simulates a specified number of concurrent users issuing transactions with a recommended mix (45\% new order transactions, 43\% order status, and 4\% each of other transactions).
Each run lasts 5 minutes preceded by a warm-up period of 2 minutes. 
We use the number of transactions per minute (TPM), and the average latency of transactions (in milliseconds) as measures for throughput and latency. 

The \performance\ database maintains employee performance data for one year (approximately 50 GB). 
The database application consists of 104 stored procedures of varying complexity (from 100 - 2500 LOC), and 69 views. 
The workload consists of a mix of 7 complex workflows, where each workflow invokes several stored procedures. 
We choose this application because it has strong security requirements but much lower throughput and latency requirements than TPC-C, and therefore serves as an interesting design point. 



\subsection{TPC-C}
For evaluating the security performance trade-off in TPC-C, we defined three progressively weaker encryption policies. 
The policy \strongp\ uses randomized encryption for all columns in the customer table containing personally identifying information.
\securesql's type system certifies that this policy does not contain any implicit or explicit insecure information flows. 
However, the policy does not utilize indexes or permit server side computation on these columns. 
The policy \deterministicp\ weakens \strongp\ by using deterministic encryption for customer last name and credit status, and leaving the column {\tt C\_DATA} in cleartext.
This policy has an implicit flow from the column {\tt C\_CREDIT} to {\tt C\_DATA}, but is able to use indexes on these columns. 
The third policy \phep\ is similar to \deterministicp\ except it uses Paillier encryption for customer's account balance. 
Our baseline is a configuration (\cleartextp) where all columns are cleartext, and the application runs entirely on the server. 
The compiled stored procedures for all policies are available at ~\cite{securesql15supplement}.
\\

\begin{table}[t]
\footnotesize
\centering
\begin{tabular}{|r|r|r|r|r|r|r|}
\hline 
\multicolumn{1}{|c|}{\textbf{Policy}} & \multicolumn{5}{|c|}{\textbf{\# Coercions}} & \multicolumn{1}{|c|}{\textbf{\# DT}}\\
\hline
& \textit{DEL} & \textit{NEWORD} & \textit{OSTAT} & \textit{PAY} & \textit{SLEV} & \\ 
\hline 
\textit{\strongp} & 3 & 3 & 8 & 44 & 0 & 4 \\ 
\hline 
\textit{\deterministicp} & 3 & 3 & 4 & 24 & 0 & 2 \\ 
\hline 
\textit{\phep} & 0 & 0 & 2 & 18 & 0 & 0 \\ 
\hline 
\end{tabular} 
\caption{\label{tab:tpcc-stats}
Number of coercions and the number of procedures requiring distributed transactions in the TPC-C benchmark for different encryption policies.} 
\end{table}

\noindent\textbf{Static properties.}
We measured two properties of a partitioned application that serve as good indicators of performance i.e. the number of coercions, and the number of distributed transactions (Table~\ref{tab:tpcc-stats}).
With policy \strongp\, 4 out of 5 stored procedures require coercions, with the procedure {\it Payment} requiring as many as 44 coercions, illustrating the complexity of rewriting. 
Both the number of coercions and the number of procedures that require distributed transactions reduces as the policy is weakened.
With policy \phep, all computation (with the exception of encryption \& decryption routines, and simple scalar operations) is offloaded to the server, and distributed transactions are eliminated altogether.
We also observe that encryption increases the space overheads of the database by roughly 45\% (with negligible variance across policies).\\


\begin{figure}[t]
\subfloat{}{}{\includegraphics[clip=true,trim=2cm 7.5cm 2cm 9cm,width=1.65in]{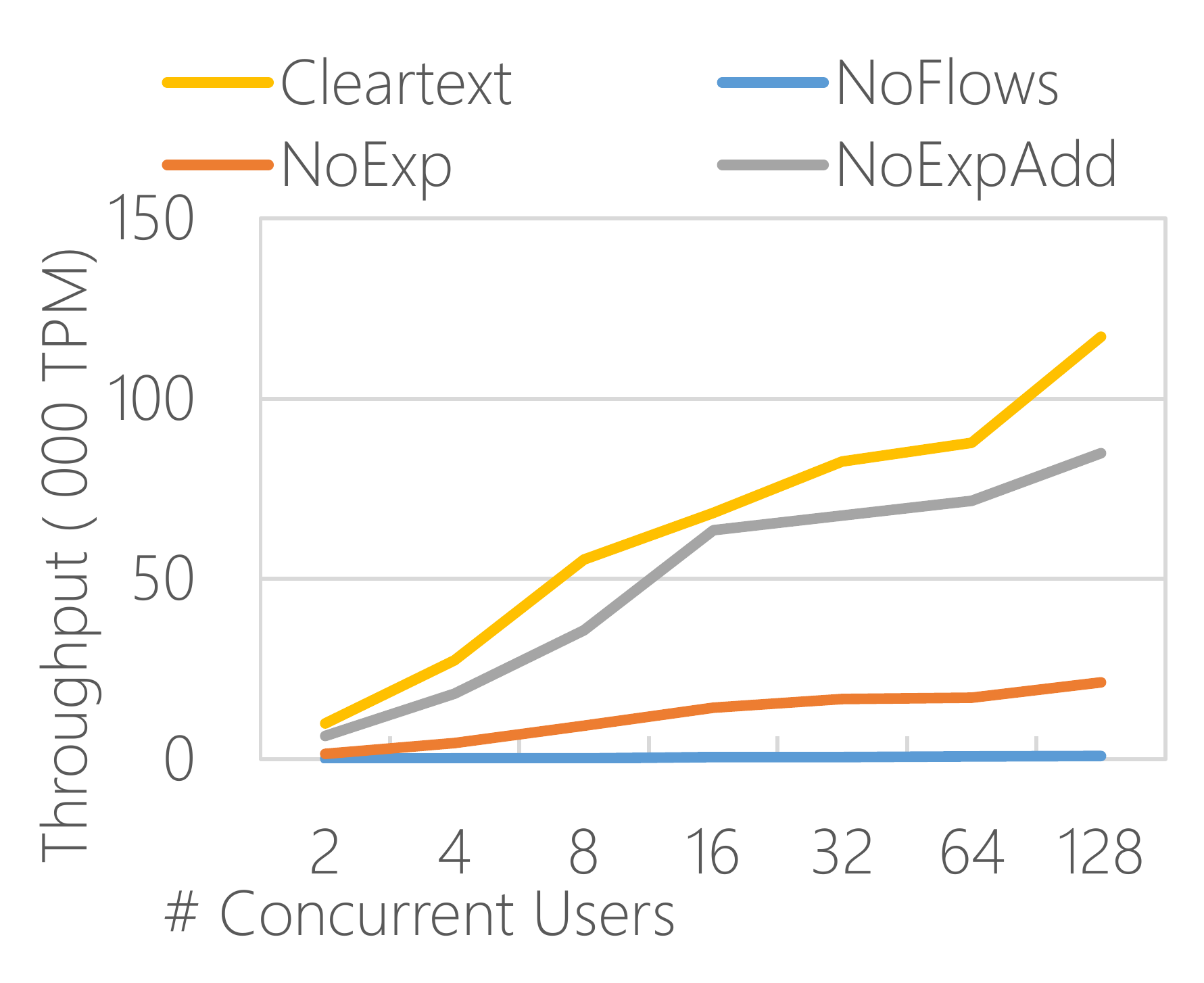}}
\subfloat{}{}{\includegraphics[clip=true,trim=2cm 7.5cm 2cm 9cm,width=1.65in]{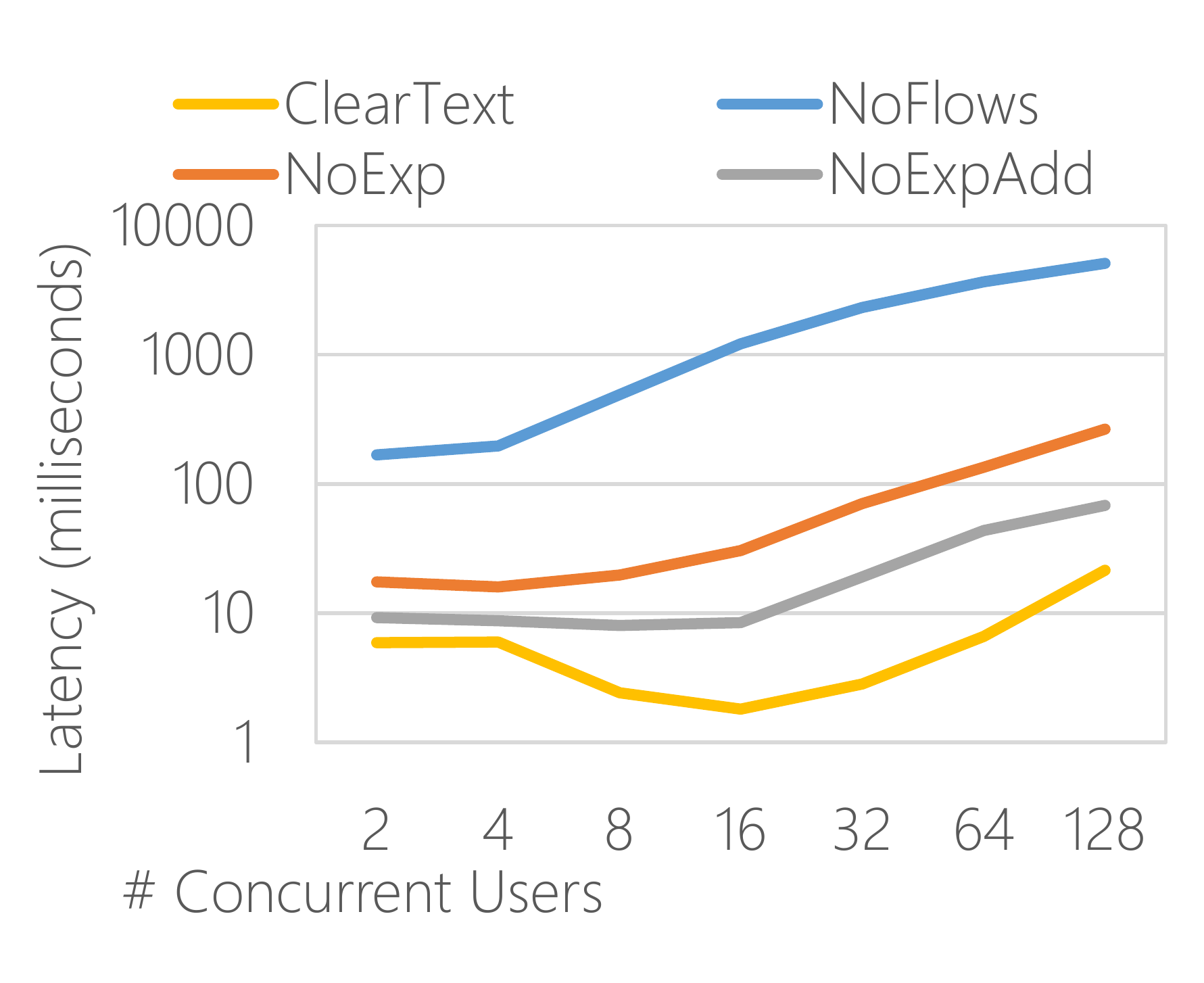}}
\caption{\label{fig:tpcc-overheads-policy}
Throughput and latency of TPCC with varying load and encryption policies.}
\end{figure}

\noindent\textbf{Performance Overheads.}
Figure~\ref{fig:tpcc-overheads-policy} shows the average throughput and latency of the partitioned TPC-C with varying number of users.
As one might expect, throughput scales almost linearly with load in the baseline configuration with all columns in cleartext. 
The throughput with the policy \phep\ closely matches the baseline, except at high load where the throughout is lower by 28\%. 
The discontinuity at 16 users is explained by the number of cores available on the server. 
Further analysis shows that the lower throughput is due to encryption and decryption operations on the trusted client. 
Partitioning has a larger effect on latency for policy \phep\ -- average latency increases from 1.8ms to 8.4ms with 16 concurrent users. 
This is due to at least one additional hop between application and the database server, and the latency of encryption and decryption operations. 

Both throughput and latency drop significantly for policies \strongp\ and \deterministicp\ (with a maximum of 1000 and 19800 TPM respectively).
The drop is performance with \strongp\ is not be entirely unexpected.
In this configuration, most of the query processing and control flow remains on the client.
An analysis of the most time consuming queries using SQL profiler reveals that cursors are the worst affected, followed by queries that filter on encrypted columns. 
Since these queries now execute on the client, performance suffers due to the inability to use indexes, and the additional data transfers to the client.
However, the drop in performance with \deterministicp\ is more surprising. 

\begin{figure}[t]
\subfloat{}{}{\includegraphics[clip=true,trim=2cm 7.5cm 2cm 9.5cm,width=1.65in]{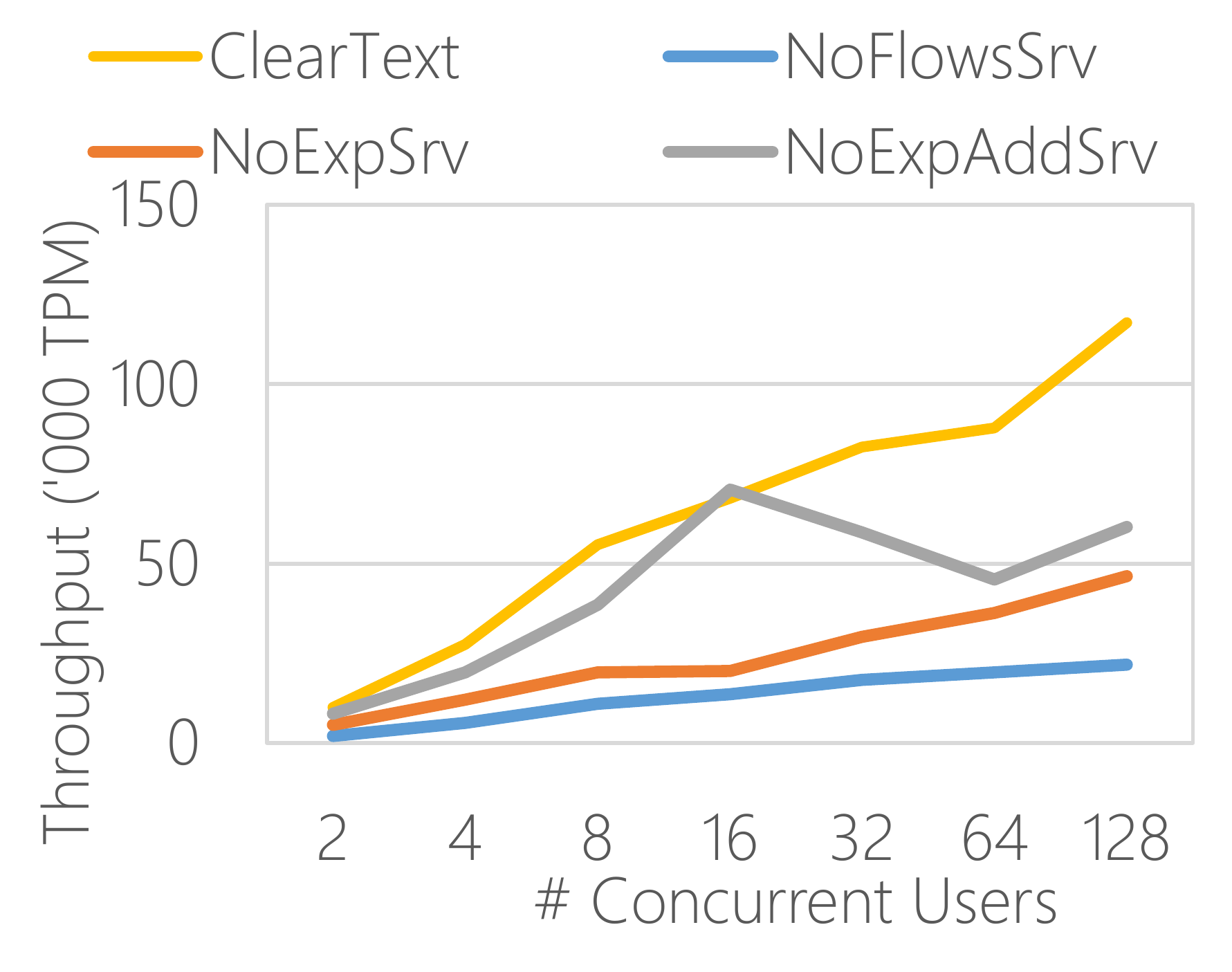}}
\subfloat{}{}{\includegraphics[clip=true,trim=2cm 7.5cm 2cm 9.5cm,width=1.65in]{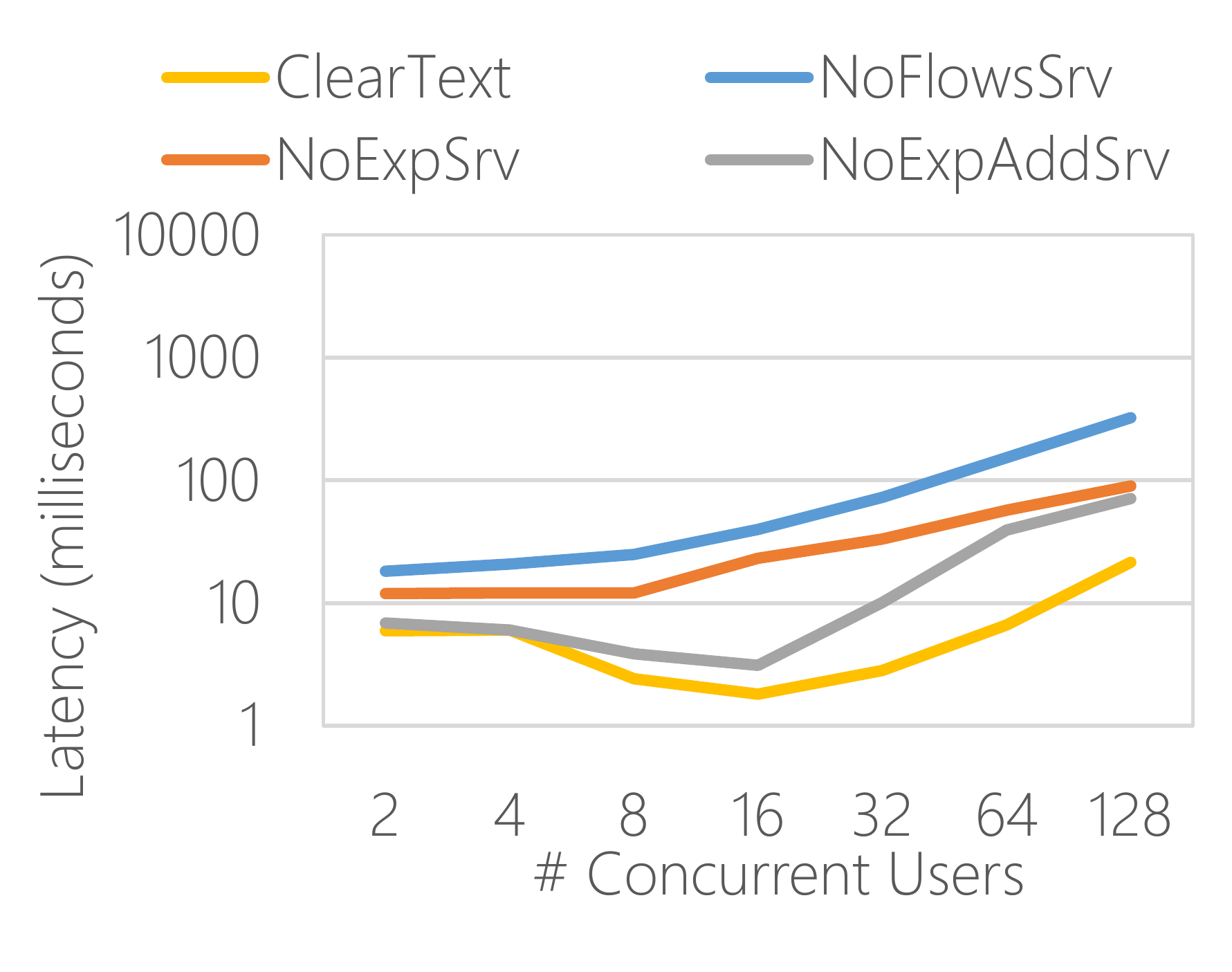}}
\caption{\label{fig:tpcc-overheads-micro}
Throughput and latency of TPCC for configurations where the server has access to keys.}
\end{figure}

We performed a set of experiments to isolate the cause of loss in performance.
We considered three additional configurations \posrv, \ptsrv\ and \pthsrv\ where the database is encrypted with policies \strongp\, \deterministicp\ and \phep\ respectively, and the database server has access to encryption keys. 
Therefore, both the server and trusted client components of the application can be deployed on the server. 
These configurations are not secure -- they simply eliminate overheads of the network. 

Figure~\ref{fig:tpcc-overheads-micro} shows the throughput and latency for these configurations.
The throughout and latency of \ptsrv\ are significantly better than \deterministicp. 
Both through and latency improve by a factor of 5, with latency within 10\% of \phep\ on average. 
Throughput is however lower than \phep\ by a factor of 3. 
This suggests that most of the overheads (if not all) can be attributed to the cost of distributed query processing. 
Recall that in \deterministicp, the column {\tt C\_BALANCE} in the customer table is encrypted using randomized encryption. 
Therefore, operations on this column are executed in the trusted client component. 
This requires additional round trips and use of distributed transactions in two stored procedures ({\tt DEL} and {\tt PAY}), which hurts throughput. 

Also observe the performance of \pthsrv\ reduces significantly at high loads. 
Further analysis of CPU usage reveals that the loss in performance is due to encryption and decryption routines, which contend for CPU cycles with the rest of the application. 
Contrast this with \phep\ where the load is shared between the client and the server.\\

\noindent\textbf{Optimizations.}
We also measured the performance of TPC-C for each of the encryption policies with the optimizations described in Section~\ref{sec:partitioning} disabled. 
In this mode, the performance of configurations \deterministicp\ and \phep\ drops significantly, with throughput even lower than \strongp. 
We also observed that no single optimization when enabled in isolation improved performance significantly. 
Performance improved only when these optimizations are enabled together. 
Therefore, optimizations play a big role in realizing the full benefits of a weaker encryption policy.


In summary, the experiments suggest that \securesql\ can ensure the absence of explicit flows even in a performance sensitive application like TPC-C with reasonably low overheads. 
However, overheads can he high if the encryption policy is not carefully chosen to maximize the use of existing PHE schemes, or protection from implicit flows is desired.
\\

\begin{figure}
\centering
\includegraphics[clip=true,trim=2cm 9.7cm 2cm 11.4cm,width=2.5in]{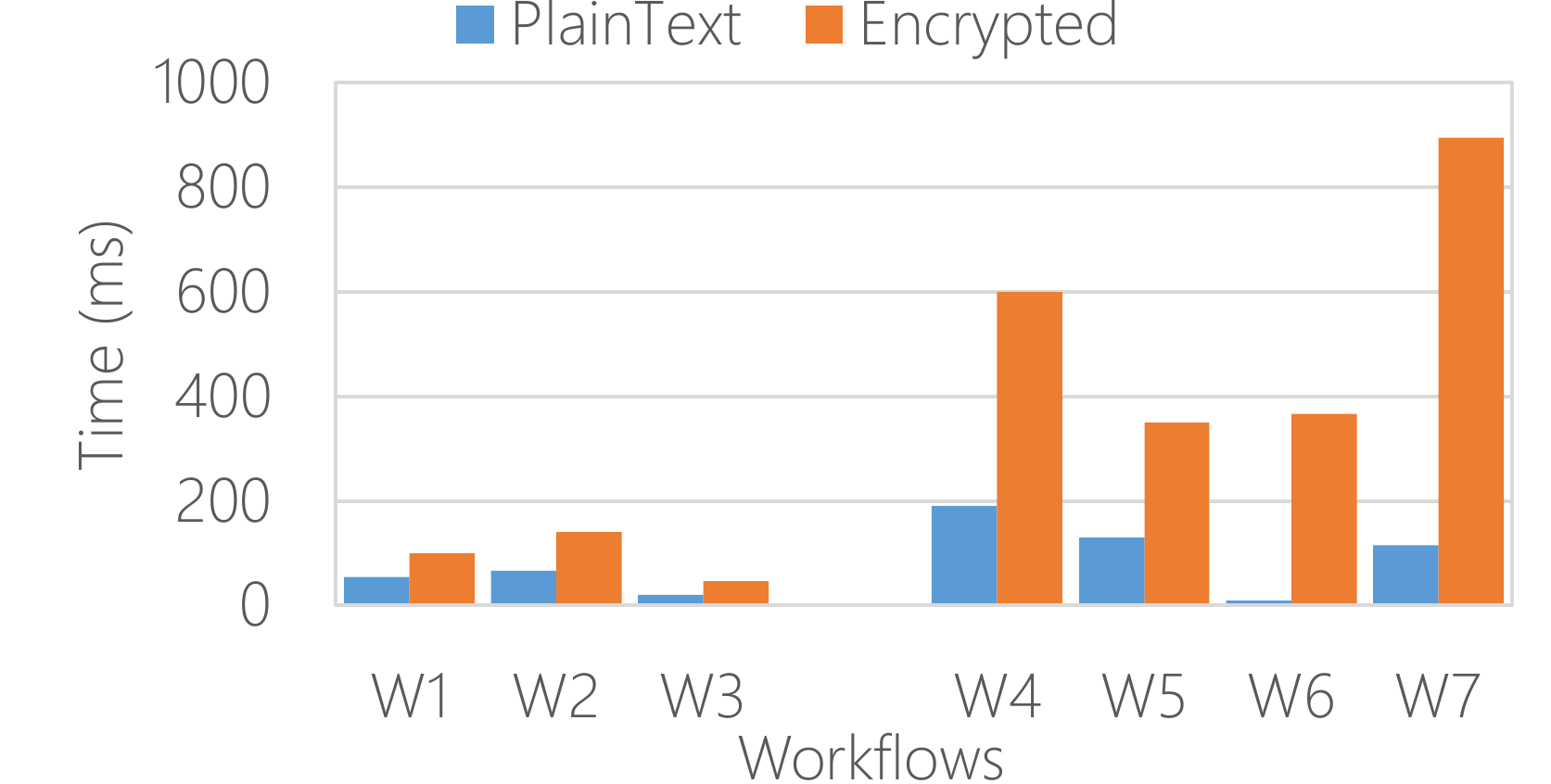}
\caption{\label{fig:pam-latency}Average latency of workflows with and without encryption.}
\vspace{-0.1in}
\end{figure}

\subsection{Employee performance evaluation}
For this application, we use a strong encryption policy where all personally identifiable data and data related to an employee's performance is encrypted using randomized encryption. 
\securesql's type system is able to verify that with this policy, the partitioned application has no implicit or explicit flows. 
Even with this policy, we find that over 80\% of the stored procedures and views do not require any coercions. 
However, there are procedures where the compiler introduces up to 46 coercions. 
Figure~\ref{fig:pam-latency} shows the average latency of 7 main workflows.
We compare the latency with a configuration where the database is in cleartext. 
The three most critical workflows in this application, W1, W2 and W3 are invoked when employees lookup their appraisals. 
These workflows are are dominated by lookups and complex joins on cleartext columns and moderately impacted by partitioning. 
The latency of workflows W4-W7 increase significantly after encryption. 
This is due to the presence of calls to string manipulation routines on encrypted data within the scope of distributed transactions, and joins over strongly encrypted columns which cannot be offloaded to the server. 
We also tested this application through its web based client -- on the whole usability of the application is not affected by partitioning.  
\section{Conclusions and Future work}
\label{sec:conclusions}
Encrypted databases are a first step towards the goal of applications which can "compute" on encrypted data and offer strong end-to-end confidentiality guarantee. 
This paper takes a step towards defining strong security properties for encrypted databases. 
We show that encrypted databases can support real-world workloads with reasonable security and performance. 
However, ensuring strong information flow security with software-only trusted clients in performance sensitive applications is challenging. 
Encrypted databases with trusted hardware~\cite{arasu2013orthogonal,mckeen2011innovative} may be able to offer both information flow security and robust performance. 
From a security perspective, formally stating the threat model and proving that \securesql's type system ensures confidentiality is an open problem. 

\vspace{-0.03in}
\section{Related Work}
\paragraph{Trusted clients.}
The trusted client model was first proposed in ~\cite{hacigumus02sqlencrypted}. 
Their system does not exploit partially homomorphic encryption. 
Instead, the system maintains additional indexes on encrypted columns, which permits some filters to be pushed to the server. 
They present an algebraic framework for partitioning individual queries along with a set of heuristics for maximizing the amount of computation pushed to the server - which requires changes to the underlying query optimizer. 
In contrast, we handle a richer set of abstractions (views/stored procedures) and require no changes to the underlying database infrastructure.

CryptDB~\cite{popa2011cryptdb}, Monomi~\cite{tu2013processing} continue the line of work in~\cite{hacigumus02sqlencrypted} by using partial homomorphic encryption schemes. 
They differ in the amount of computation that is allowed in the trusted client - CryptDB uses a web proxy and is not a general purpose system. 
Monomi permits arbitrary residual computation on the client, but is geared towards analytical workloads. 
The main focus in both these systems is individual queries - in this paper we study abstractions such as stored procedures, views, and transactions in addition to ad-hoc queries. 
Interestingly, these abstractions fundamentally change the security model - in particular, stored procedures introduce information flow between columns.
Any system that ignores these information flows can potentially leak the contents of encrypted columns to a passive adversary. 
Our compiler uses a type system that statically checks for insecure information flows and rewrites stored procedures. 
Finally, we evaluate our system using complete benchmarks (as opposed to traces used in~\cite{popa2011cryptdb}). 
Our evaluation offers a more realistic picture of the performance of a fully general trusted client based database architecture for transactional workloads. \\

\paragraph{Untrusted clients.}
Chong et al~\cite{chong2007secure} propose a system for automatically partitioning web applications. 
In their security model, the client is considered untrusted.
The objective of partitioning the web application is to ensure that sensitive data does not flow to the client. 
Our security property is a variant of the information flow security tailored for partially homomorphic encryption schemes, which their system does not use.\\

\paragraph{Hardware based security.}
Another approach for securely processing query on untrusted platform using dedicated trusted hardware such as FPGAs and Intel SGX~\cite{mckeen2011innovative}.
TrustedDB~\cite{bajaj2011trusteddb} combines a secure co-processor and a commodity server. 
It runs a lightweight on the secure co-processor and a full-fledged database on the server.
Query processing is distributed between two databases - encrypted data is processed on the co-processor and cleartext data is processed using a commodity database. 
Cipherbase~\cite{arasu2013orthogonal} also relies on secure hardware (in the form of an FPGA device) to process encrypted data. 
However, unlike TrustedDB, the database is more tightly coupled with the hardware. 
The database performs as much computation as possible while delegating computation that requires sensitive data in cleartext to secure hardware. 
Haven~\cite{baumann2014shielding} is a system for running unmodified applications on an untrusted platform using SGX. 
While Haven can isolate the entire database from the platform, security is predicated on a large trusted computing base, which includes the database server and the guest operating system. 
Haven also does not offer strong information flow properties. 
Compared to an on-premises trusted client, a hardware based approach can potentially yield better performance due to lower round trip latencies.
A key deficiency of hardware based approaches is that they are intrusive and cannot be implemented on top of existing infrastructure. 
Another shortcoming is that these systems focus on data confidentiality but do not consider leaks due to information flow. 
It should be possible to extend our type system and partitioning algorithm to target trusted hardware instead of a trusted client and provide stronger security guarantees. 
\\

\bibliographystyle{plain}
\bibliography{main}
\end{document}


\title{Information Flows in Encrypted Databases : Supplement}
\date{}
\maketitle
\thispagestyle{empty}
\section{Motivating example}
\label{sec:overview}

Figure~\ref{fig:tpcc-payment-implicit} shows the partitioning that contains no explicit or implicit flows. 
This partitioning requires the column {\textit C\_DATA} to be encrypted, and consequently incurs an additional round trip to the client because the string concatenation operation must evaluated on the client. 

\begin{figure}[!h]
\lstset{basicstyle=\ttfamily\scriptsize,tabsize=1,numbers=left,language=sql}
\begin{lstlisting}
--Server
CREATE PROCEDURE [dbo].[__Closure]
@c_w_id INT, @h_amount VARBINARY(2048), @c_last VARBINARY(256)
AS BEGIN
	SELECT @pubkey = ...
	UPDATE dbo.CUSTOMER
	SET @c_id = C_ID,
		@c_first =  C_FIRST, 
		@c_credit = C_CREDIT,
		@c_balance = C_BALANCE = 
			dbo.PaillierAdd(C_BALANCE, @h_amount, @pubkey)
	WHERE CUSTOMER.C_W_ID = @c_w_id
		AND CUSTOMER.C_LAST = @c_last;

	INSERT dbo.HISTORY (H_C_ID, H_C_BALANCE)
	VALUES (@c_id, @c_balance)

	SELECT @c_id AS N'@c_id', 
		@c_first AS N'@c_first',
		@c_last AS N'@c_last', 
		@c_credit AS N'@c_credit',
		@c_balance AS N'@c_balance'
END

-- Shell
CREATE PROCEDURE [dbo].[PAYMENT]
@c_w_id INT, @h_amount NUMERIC(6,2), @c_last CHAR(16)
AS BEGIN
	SELECT @key = ... // private key 
	SELECT @pubkey = ...	// public key for paillier
	SELECT 
		@enc_h_amount = 
			dbo.AEncrypt(@h_amount, @key, @pubkey),
		@enc_c_last = dbo.DEncrypt(@c_last, @key),
	
	BEGIN TRANSACTION
		EXEC [SERVER].[tpcc].dbo.__Closure1 
			@enc_c_w_id, @c_d_id,
			@enc_c_amount, @enc_c_last,
			out @c_id, out @c_first, out @c_last, 
			out @c_balance, out @c_credit	

		if (@c_credit = 0x002057E9A8865AAA7D59DA69AD...)
			UPDATE [SERVER].dbo.CUSTOMER
			SET C_DATA = dbo.REncrypt(@h_amount + C_DATA)
			WHERE CUSTOMER.C_W_ID = @c_w_id
				AND CUSTOMER.C_LAST = @enc_c_last;

		SELECT @c_id AS @c_id,
			dbo.RDecrypt(@c_first, @key) AS @c_first,
			dbo.DDecrypt(@c_last, @key) AS @c_last,
			dbo.DDecrypt(@c_credit, @key) AS @c_credit,
			dbo.ADecrypt(@c_balance, @key, @pubkey) AS @c_balance,
	COMMIT TRANSACTION
END
\end{lstlisting}
\caption{
\label{fig:tpcc-payment-implicit}
A T-SQL procedure derived from TPC-C}
\end{figure}
\begin{figure*}[!t]
\small
\begin{mathpar}
\inferrule* [Lab={\tiny [CONST]}]
{\beta=\fresh}
{\{\spt \subtyperel \beta\},\{\val : \spt\}\satisfies \val : \beta}
\and
\inferrule* [Lab={\tiny [VAR]}]
{\alpha,\beta=\fresh}
{\{\alpha \subtyperel \beta\},\{\var : \alpha\}\satisfies \var : \beta}
\and
\inferrule* [Lab={\tiny [COLUMN]}]
{\alpha=\policy(\tablexpr{}, \column{})\judgespace\beta=\fresh}
{\{\alpha \subtyperel \beta\},\{\var : \alpha\}\satisfies \tablexpr{}.\column\ : \beta}
\and
\inferrule* [Lab={\tiny [REC]}]
{\forall i \in (1..n)\ \judgeexpr{i}\judgespace\subs=\unify(\{\alpha,\beta \mid \var : \alpha \in \typeenv_{j} \wedge \var : \beta \in \typeenv_{k}\}}
{\cup_{i=1}^{n}\subs(\constraints_{i}),\cup_{i=1}^{n}\subs(\typeenv_{i})\satisfies\rec : \ensuremath{[\subs(\beta_{1}),\dots,\subs(\beta_{n})]}}
\and
\inferrule* [Lab={\tiny [EQUALS]}]
{\judgeexpr{1}\judgespace\judgeexpr{2}\judgespace\beta=\fresh \subs=\unify(\{\alpha,\beta \mid \var : \alpha \in \typeenv_{1} \wedge \var : \beta \in \typeenv_{2}\} \cup \{\beta_1, \beta_2\})}
{\subs(\constraints_{1})\cup\subs(\constraints_{2}) \cup \{\beta\in [\spt,\oppt,\dpt,\ndpt]\} \cup\{\beta_{1}\in [\sope,\sopel,\sde,\sdel,\spt,\oppt,\dpt,\ndpt],\\\\\subs({\beta_1})\subtyperel\beta\},\subs(\typeenv_{1})\cup\subs(\typeenv_{2})\satisfies\exprsub{1} = \exprsub{2} : \beta}
\and
\inferrule* [Lab={\tiny [COMP]}]
{\judgeexpr{1}\judgespace\judgeexpr{2}\judgespace\beta=\fresh\judgespace\subs=\unify(\{\alpha,\beta \mid \var : \alpha \in \typeenv_{1} \wedge \var : \beta \in \typeenv_{2}\} \cup \{\beta_1, \beta_2\})}
{\subs(\constraints_{1})\cup\subs(\constraints_{2}) \cup \{\beta\in [\spt,\oppt,\dpt,\ndpt]\} \cup\{\beta_{1}\in [\sope,\sopel,\spt,\oppt,\dpt,\ndpt],\\\\\subs({\beta_1})\subtyperel\beta\},\subs(\typeenv_{1})\cup\subs(\typeenv_{2})\satisfies\exprsub{1} < \exprsub{2} : \beta}
\and
\inferrule* [Lab={\tiny [ADD]}]
{\judgeexpr{1}\judgespace\judgeexpr{2}\judgespace\beta=\fresh\judgespace\subs=\unify(\{\alpha,\beta \mid \var : \alpha \in \typeenv_{1} \wedge \var : \beta \in \typeenv_{2}\} \cup \{\beta_1, \beta_2\})}
{\subs(\constraints_{1})\cup\subs(\constraints_{2})\cup\{\beta_1 \in [\sade,\sadel,\spt,\oppt,\dpt,\ndpt],\subs({\beta_1})\subtyperel\beta\},\subs(\typeenv_{1})\cup\subs(\typeenv_{2})\satisfies\exprsub{1} + \exprsub{2} : \beta}
\and
\inferrule* [Lab={\tiny [APPLY]}]
{\judgeexpr{1}\judgespace\judgeexpr{2}\judgespace\tau,\mu=\fresh\judgespace\subs=\unify(\{\alpha,\beta \mid \var : \alpha \in \typeenv_{1} \wedge \var : \beta \in \typeenv_{2}\} \cup \{\beta_1, \beta_2\rightarrow\tau\})}
{\subs(\constraints_{1})\cup\subs(\constraints_{2})\cup\{\tau\subtyperel\mu\},\subs(\typeenv_{1})\cup\subs(\typeenv_{2})\satisfies\func\expr : \mu}
\and
\inferrule* [Lab={\tiny [ABS]}]
{\constraints,\typeenv\cup \{\var : \alpha\}\satisfies\expr : \beta\judgespace\tau=\fresh}
{\constraints\cup\{\alpha\rightarrow\beta\subtyperel\tau\},\typeenv\satisfies\func(\var).\expr : \tau}
\and
\inferrule* [Lab={\tiny [SELECT]}]
{\judgeexpr{1}\judgespace\judgeexpr{2}\judgespace\beta=\fresh\judgeexpr\subs=\unify(\{\alpha,\beta \mid \var : \alpha \in \typeenv_{1} \wedge \var : \beta \in \typeenv_{2}\}\cup\{\beta_1,\pt\})}
{\subs(\constraints_{1})\cup\subs(\constraints_{2})\cup\{\subs(\beta_{2})\subtyperel\beta\},\subs(\typeenv_{1})\cup\subs(\typeenv_{2})\satisfies\selectexpr:\beta}
\and
\inferrule* [Lab={\tiny [PROJECT]}]
{\forall{i},\ \beta_{i}=\policy(\tabl{},\column{i})\judgespace\beta=\fresh}
{\{[\beta_1,\dots,\beta_n]\subtyperel\beta\},\{\}\satisfies: \projectexpr:\beta}
\and
\inferrule* [Lab={\tiny [UNION]}]
{\judgeexpr{1}\judgespace\judgeexpr{2}\judgespace\subs=\unify(\{\alpha,\beta \mid \var : \alpha \in \typeenv_{1} \wedge \var : \beta \in \typeenv_{2}\}\cup\{\beta_1,\beta_2\})\judgespace\beta=\fresh}
{\subs(\constraints_{1})\cup\subs(\constraints_{2})\cup\{,\subs({\beta_1})\subtyperel\beta\},\subs(\typeenv_{1})\cup\subs(\typeenv_{2})\satisfies\unionexpr:\beta}
\and
\inferrule* [Lab={\tiny [DIFF]}]
{\judgeexpr{1}\judgespace\judgeexpr{2}\judgespace\subs=\unify(\{\alpha,\beta \mid \var : \alpha \in \typeenv_{1} \wedge \var : \beta \in \typeenv_{2}\}\cup\{\beta_1,\beta_2\})\judgespace\beta=\fresh}
{\subs(\constraints_{1})\cup\subs(\constraints_{2})\cup\{,\subs({\beta_1})\subtyperel\beta\},\subs(\typeenv_{1})\cup\subs(\typeenv_{2})\satisfies\diffexpr:\beta}
\and
\inferrule* [Lab={\tiny [PRODUCT]}]
{\forall{i}\in (1,n),\ \beta_{i}=\policy(\tabl{1},\column{i})\judgespace\forall{j}\in(1,m),\ \tau_{j}=\policy(\tabl{2},\column{j})\judgespace\beta=\fresh}
{\subs(\constraints_{1})\cup\subs(\constraints_{2})\cup\{[\beta_{1},\dots,\beta_{n},\tau_{1},\dots,\tau_{m}]\subtyperel\beta\},\subs(\typeenv_{1})\cup\subs(\typeenv_{2})\satisfies\productexpr:\beta}
\end{mathpar}
\caption{
\label{fig:type-inference}
Type inference rules for expressions in \LSQL}
\end{figure*}

\begin{figure*}[!t]
\small
\begin{mathpar}
\and
\inferrule* [Lab={\tiny [ASSIGN]}]
{\constraints_{1},\typeenv_{1}\cup \{\exprsub{1} : \alpha_{1}\}\satisfies\exprsub{1} : \beta_{1}\judgespace\judgeexpr{2}\judgespace\subs=\unify(\{\alpha,\beta \mid \var : \alpha \in \typeenv_{1} \wedge \var : \beta \in \typeenv_{2}\}\cup\{\alpha_{1},\beta_{1}\})}
{\subs(\constraints_{1})\cup\subs(\constraints_{2})\cup\{\subs(\beta_{2})\subtyperel\subs(\beta_{1})\},\subs(\typeenv_{1})\cup\subs(\typeenv_{2}),\subs(\beta_{1})\satisfies\exprsub{1}:=\exprsub{2}}
\and
\inferrule* [Lab={\tiny [INSERT]}]
{\forall{i},\ \beta_{i}=\lvaltype({\policy(\tablexpr{},\column{i})})\judgespace\judgeexpr{}\judgespace\gamma=\fresh}
{\constraints\cup\{\beta\subtyperel[\beta_1,\dots,\beta_n]\cup\{\forall{i}\in[1,\dots,n],\subs(\beta_{i})\subtyperel\gamma\},\typeenv,\gamma\satisfies: \insertexpr}
\and
\inferrule* [Lab={\tiny [DELETE]}]
{\forall{i},\ \beta_{i}=\lvaltype({\policy(\tablexpr{},\column{i})})\judgespace\judgeexpr{}\judgespace\gamma=\fresh}
{\constraints\cup\{\beta\subtyperel[\beta_1,\dots,\beta_n]\cup\{\forall{i}\in[1,\dots,n],\subs(\beta_{i})\subtyperel\gamma\},\typeenv,\gamma\satisfies: \deleteexpr}
\and
\inferrule* [Lab={\tiny [UPDATE]}]
{\forall{i},\ \beta_{i}=\lvaltype({\policy(\tablexpr{},\column{i})})\judgespace\judgeexpr{}\judgespace\gamma=\fresh}
{\constraints\cup\{\beta\subtyperel[\beta_1,\dots,\beta_n]\cup\{\forall{i}\in[1,\dots,n],\subs(\beta_{i})\subtyperel\gamma\},\typeenv,\gamma\satisfies: \updateexpr}
\and
\inferrule* [Lab={\tiny [SELECT]}]
{\judgeexpr{}\judgespace\forall{i}\in[1,\dots,m],\ \beta_{i}=\project(\beta, \name{i})\judgespace\gamma=\fresh}
{\{\constraints\cup\{\forall{i}\in[1,\dots,n],\subs(\beta_{i})\subtyperel\gamma\},\typeenv,\gamma\satisfies: \selectstmt}
\and
\inferrule* [Lab={\tiny [IF]}]
{\judgeexpr{1}\judgespace\judgestmt{1}\judgespace\judgestmt{2}\judgespace\subs=\unify(\{\alpha,\beta \mid \var : \alpha \in \typeenv_{i} \wedge \var : \beta \in \typeenv_{j}\} \cup \{\gamma_{1}, \gamma_{2}\})}
{\subs(\constraints_{1})\cup\subs(\constraints_{2})\cup\subs(\constraints_{3})\cup\{\subs(\beta_{1})\subtyperel\subs(\gamma_{1})\},\subs(\typeenv_{1})\cup\subs(\typeenv_{2})\cup\subs(\typeenv_{3}),\subs(\gamma_{1})\satisfies\ifexpr}
\end{mathpar}
\caption{
\label{fig:type-inference-statements}
Type inference rules for statements in \LSQL}
\end{figure*}

\section{Type inference}
The type inference algorithm is described as a set of type rules.